\documentclass[12pt]{iopart}

\usepackage{dcolumn}
\usepackage{comment}
\usepackage{hyperref}
\usepackage{epstopdf}% To incorporate .eps illustrations using PDFLaTeX, etc.
\usepackage{placeins}
\usepackage{color}

\usepackage[separate-uncertainty=true]{siunitx}

\usepackage[utf8]{inputenc}
\usepackage[T1]{fontenc}
\usepackage{graphicx}
\usepackage{cite}

\expandafter\let\csname equation*\endcsname\relax
\expandafter\let\csname endequation*\endcsname\relax

\usepackage{amssymb}
\usepackage{amsmath} %right equation numbering + mathenvironments
\usepackage{mathtools}
\usepackage{bm} 

\usepackage{xfrac} %%\sfrac for 1/2 type fractions

\newcommand{\tx}[1]{\textnormal{#1}}
\def\@fnsymbol#1{\ensuremath{\ifcase#1\or *\or \dagger\or \ddagger\or
   \mathsection\or \mathparagraph\or \|\or **\or \dagger\dagger
   \or \ddagger\ddagger \else\@ctrerr\fi}}
\newcommand{\ssym}[1]{^{\@fnsymbol{#1}}}

\begin{document}

\def\BY{\begin{eqnarray}}
\def\EY{\end{eqnarray}}
\def\BI{\begin{itemize}}
\def\EI{\end{itemize}}
\def\L{\label}
\def\nn{\nonumber}
\def\({\left (}
\def\){\right)}
\def\[{\left [}
\def\]{\right]}
\def\<{\langle}
\def\>{\rangle}
\def\BA{\begin{array}}
\def\EA{\end{array}}
\def\dsp{\displaystyle}
\def\ds{\displaystyle}
\def\k{\kappa}
\def\dd{\delta}
\def\D{\Delta}
\def\w{\omega}
\def\W{\Omega}
\def\v{\nu}
\def\a{\alpha}
\def\b{\beta}
\def\e{\varepsilon}
\def\exp{\text{e}}
\def\d{\partial}
\def\g{\gamma} %response rate
\def\G{\Gamma}
\def\tt{\theta}
\def\t{\tau}
\def\s{\sigma}
\def\+{\dag}
\def\8{\infty}
\def\x{\xi}
\def\m{\mu}
\def\l{\lambda}
\def\ii{\text{i}}
\def\={\approx}
\def\xc{\frac{2x}{c}}
\def\->{\rightarrow}
\def\r{\vec{r}}
\def\k{\vec{k}}
\def\sinc{\mathrm{sinc}}
\def\xx{\textbf{x}}
\def\yy{\textbf{y}}
\def\qq{\textbf{q}}
\def\rr{\boldsymbol{\rho}}
\newcommand{\ud}{\,\mathrm{d}} % For upright d in the integrals
\def\out{|\textrm{out}\rangle}
\def\inn{|\textrm{in}\rangle}
\def\sqz{|\textrm{sqz}\rangle}
\def\A{\text{A}}
\def\B{\text{B}}
\def\AB{\text{AB}}
\def\where{\text{where:}}
\def\F{{\cal F}}
\def\A{{\cal A}}
\def\inn{\text{in}}
\def\o{\text{out}}
\def\bU{\bold{U}}
\def\bV{\bold{V}}
\def\bu{\bold{u}}
\def\bv{\bold{v}}
\def\si{\text{s,i}}
\def\s{\text{s}}
\def\i{\text{i}}

\title{
Off-resonant emission of photon pairs in nonlinear optical cavities
}
%\pdfoutput=1
\author{
Valentin A. Averchenko$^{1,*}$, Gerhard Schunk$^{1,2,3,*}$,
Michael Förtsch$^{4}$,
Martin Fischer$^{1,2}$,
Dmitry V. Strekalov$^{1,2}$,
Gerd Leuchs$^{1,2}$,
Christoph Marquardt$^{1,2}$
}
\address{$^{1}$Max Planck Institute for the Science of Light, Staudtstra{\ss}e 2, 91058 Erlangen, Germany}              
\address{$^{2}$Institute for Optics, Information and Photonics, University Erlangen-N\"{u}rnberg, Staudtstra{\ss}e 7/B2, 90158 Erlangen, Germany}
\address{$^{3}$SAOT, School in Advanced Optical Technologies, Paul-Gordan-Stra{\ss}e 6, 91052 Erlangen, Germany}
\address{$^{4}$TRUMPF GmbH + Co. KG, Johann-Mausstra{\ss}e 2,
71522 Ditzingen, Germany}
\address{$^{*}$ These authors contributed equally to this work}

%%%%%%%%%%%%%%%%%%%%%%%%%%%%%%%%%%%%%%%%%%%%%%%%%%%%%%

\begin{abstract}
Cavity-assisted spontaneous parametric down-conversion (SPDC) and spontaneous four-wave mixing (SFWM) in nonlinear optical materials are practical and versatile methods to generate narrowband time-energy entangled photon pairs.
Time-energy entangled photons with tailored spectro-temporal properties are particularly useful for efficient quantum optical interfaces.
In this work we study the generation of photon pairs in cavity-assisted SPDC and SFWM for the general case of off-resonant conversion, namely, when the frequencies of the generated photons do not match the cavity resonances.
Such a frequency mismatch in particular depends on temperature
and requires an additional control in the experiment.
First, we propose a generic model, for description of cavity-assisted SPDC and SFWM.
We show that in both processes the mismatch reduces the generation rate of photons, distorts the spectrum and the auto-correlation function of the generated fields, as well as affects the photon generation dynamics.
Second, we verify the results experimentally using parametric generation of photon pairs in a nonlinear whispering gallery mode resonator (WGMR) as an experimental platform with controlled frequency mismatch.
Our work reveals the role of the frequency mismatch in the photon generation process and shows a way to control it.
Obtained results constitute one more step in the direction of full control over the spectro-temporal properties of entangled photon pairs and the heralded generation of  single-photon pulses with a tailored temporal mode.

\end{abstract}

\maketitle 

%\tableofcontents

%%%%%%%%%%%%%%%%%%%%%%%%%%%%%%%%%%%
\section{Introduction}

Narrowband time-energy entangled photon pairs are required for fundamental studies in quantum optics as well as for the implementation of quantum enabled optical technologies \cite{Kimble2008}.
The photon pairs are particularly suited to implement efficient quantum photonic interfaces with quantum emitters \cite{Lvovsky2009, Simon2010, Specht2011, Sangouard2011, Fekete2013} and constitute an essential quantum resource to herald narrowband single photons \cite{Neergaard-Nielsen:07, Scholz2009,Spring:13} in a tailored temporal mode \cite{Averchenko2016}.
Developing methods for full control over the spectro-temporal properties of the entangled photon pairs is the next step for research in quantum optics.

\begin{figure}[htb]
  \centering
  \includegraphics[scale=0.6]{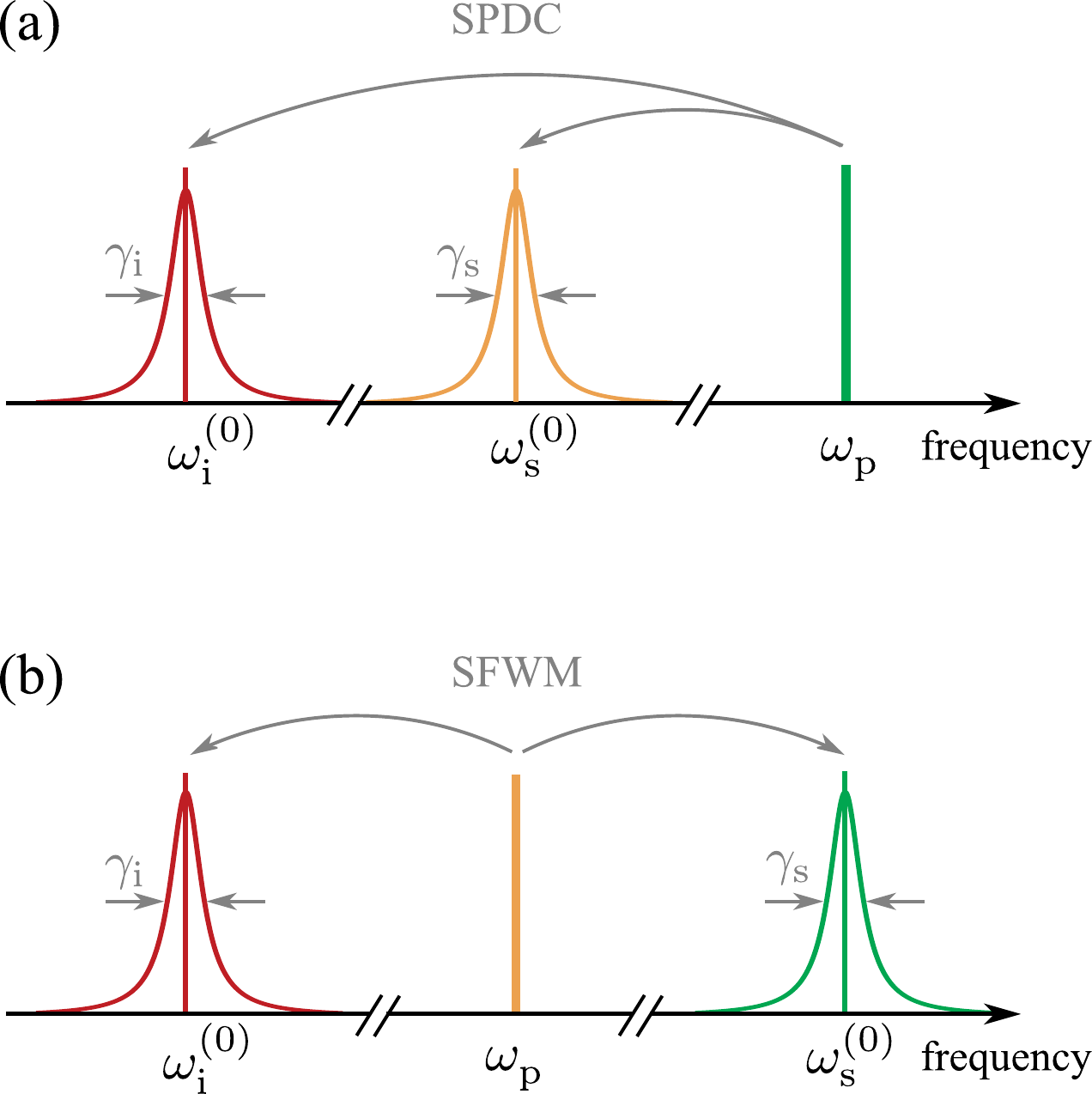} 
\caption{Generation of narrowband entangled photon pairs via a cavity-assisted spontaneous parametric down-conversion in (a) or spontaneous four-wave mixing in (b). 
Lorentzian profiles represent the cavity spectral lines with central resonance frequencies $\w_\tx{s,i}^{(0)}$ and full widths at half maximum $\g_\tx{s,i}$.
Central frequency of the pump laser is $\w_\tx{p}$.
}
\label{fig:SPDC_SFWM}
\end{figure}

A common method to generate narrowband entangled photon pairs is cavity-assisted spontaneous parametric down-conversion (SPDC). During this process single pump photons are converted into pairs of photons in a second-order nonlinear medium \cite{Ou1999}.
An optical cavity enhances the efficiency of the conversion and defines distinct narrowband modes populated by the generated photons (see Fig.~\ref{fig:SPDC_SFWM}a).
The method is low-cost and can be implemented in various designs of optical cavities  \cite{Neergaard-Nielsen:07, Fortsch2013, KHLuo2015}.
Another promising method for generation of entangled photons in the telecom wavelength range, which has been developed in the recent years, uses a spontaneous four-wave mixing (SFWM) process 
\footnote{SFWM is sometimes called spontaneous hyper-parametric conversion (see, for example, \cite{Matsko2005ReviewOA, Strekalov2016}). We adopt this terminology and refer to both processes, SPDC and SFWM, at once as "parametric conversion" throughout the paper.}
in an optical cavity to convert pairs of pump photons into pairs of correlated photons in a third-order nonlinear optical material (see Fig.~\ref{fig:SPDC_SFWM}b).
Its implementation in microring silicon resonator is promising for integrated silicon chips \cite{Grassani:15, Lu2016}.

In cavity-assisted parametric conversion  the frequency $\w_\tx{p}$ of the pump field driving the respective nonlinear process is defined by the pump laser.  The signal and idler cavity resonances $\w_\tx{s,i}^{(0)}$ are defined by geometrical/dispersion properties of the cavity and depend on the temperature.
Generally, the pump frequency may be detuned from the sum of frequencies of the signal and idler cavity resonances by a frequency mismatch $\D$:
	\begin{alignat}{2}
	&\D = \w_\tx{p}-\w_\tx{s}^{(0)} - \w_\tx{i}^{(0)} && \quad\text{(SPDC)}, 	\L{d}\\
	&\D = 2\w_\tx{p}-\w_\tx{s}^{(0)} - \w_\tx{i}^{(0)} && \quad\text{(SFWM)}\L{d1}.
	\end{alignat}
Such a frequency mismatch reduces the photon generation rate \cite{Eckardt:91} due to the off-resonant interaction of the fields. 
However, a systematic study of the frequency mismatch influence on the spectro-temporal properties of the generated photon pairs has been missing up to now.
Such studies are crucial since the spatio-temporal properties of the heralded single photons result from entanglement properties of the initial photon pairs.
Above the oscillation threshold, the frequency mismatch can be monitored via the frequency-dependent pump depletion or the conversion efficiency  \cite{Breunig2014}.
Below the threshold, there is no appreciable self-seeding of the nonlinear processes, and typically a wider spectrum of spontaneously generated photons is observed \cite{Tang1995, Zielinska2014}. 
As a result, a direct measurement and control of the frequency mismatch $\Delta$ below the oscillation threshold is experimentally challenging so far.

In this work, we study theoretically and experimentally the effect of the frequency mismatch  ((\ref{d}) and (\ref{d1})) on spectro-temporal properties of photon pairs generated in cavity-assisted parametric conversion both in the continuous and the pulsed pump regimes.
We use a simple theoretical model, which is applicable to parametric conversion in an optical cavity, both in doubly and triply resonant configurations, when the pump resonates in the optical cavity.
We analyze the generation rate, 
second-order cross- and auto-correlation functions of the photon pairs, the spectrum of photons and their temporal distribution depending on the frequency mismatch.
We experimentally confirm a number of the predicted effects using SPDC in a nonlinear triply resonant whispering-gallery mode resonator, which is a versatile source of narrowband single-mode photon pairs \cite{Fortsch2013}.
We expect that the studied effects of the frequency mismatch on the spectro-temporal properties of the entangled photon pairs are common to various implementations of cavity-assisted parametric conversion (both in doubly and triply resonant configurations) and that these effects can be observed to affect photon pairs generation in various implementations of the cavity-assisted nonlinear processes.

This work is organized as follows.
%%%%%%%
First part of the work comprises the theoretical analysis of spectro-temporal properties of narrowband photon pairs generated in cavity-assisted parametric conversion in the presence of frequency mismatch.
In section \ref{sec:model} we present a general model 
of entangled quantum state of photon pairs generated in parametric conversion.
In section \ref{sec:CW} we consider the continuous pump regime of parametric conversion and analyze the following characteristics of the photon pairs for arbitrary values of the frequency mismatch: second-order temporal cross- and auto-correlation functions, spectral densities and temporal fluxes of photons.
In section \ref{sec:pulsed} we consider the pulsed pump regime and calculate the time-dependent fluxes of generated photons depending on the frequency mismatch.

The second part presents experimental results obtained with SPDC in a monolithic nonlinear whispering-gallery mode resonator.
In section \ref{sec:scheme}, we present our experimental scheme and a characterization of our resonator in terms of the pump and the parametric bandwidths. 
In section \ref{sec:delta}, we experimentally determine the frequency mismatch $\Delta$ by measuring the signal photon count rate depending on the pump laser frequency. This evaluation corresponds in good approximation to a continuous wave pump operation. 
In section \ref{sec:exp_pulsed}, we study SPDC dynamics with nanosecond pump pulses for different values of the frequency mismatch.
Section \ref{concl} presents conclusions and outlooks.
In the \ref{appx:triply} we show the modification of the theoretical model to describe a triply resonant configuration.
In the \ref{appx:HL} we present a Heisenberg-Langevin theory for cavity-assisted parametric conversion with a time-dependent pump field.

\section{Theoretical background}\L{sec:th}

%%%%%%%%%%%%%%%%%%%%%%%%%%%%%%%%%%%
\subsection{Cavity-assisted SPDC and SFWM with a frequency mismatch}\L{sec:model}

Spontaneous parametric conversion in an optical cavity results in the generation of photon pairs in spatio-spectral optical modes, defined by the cavity.
External spectral filtering \cite{Neergaard-Nielsen:07, Haase:09}, clustering effect in a specially designed nonlinear waveguide with mirror coatings \cite{KHLuo2015}, strong second-order dispersion in a coating-free nonlinear whispering-gallery mode resonator \cite{Fortsch2013, Fortsch2014} can be employed to produce photons only in a single pair of narrowband modes.
Thus we consider only one pair of parametrically coupled modes with resonance frequencies $\w_\tx{s}^{(0)}$ and $\w_\tx{i}^{(0)}$, respectively.
We also assume that the probability to generate more than one photon pair per time interval of interest is negligible.
Therefore, we approximate the two-photon quantum state of signal and idler modes at the output of the cavity as
\footnote{
In \ref{appx:HL} we analyze the properties of the parametrically generated signal and idler fields in the Heisenberg picture. The analysis confirms our results and gives explicitly the proportionality coefficients omitted in the analysis based on the Schr{\H o}dinger picture.
}
	\begin{align}
	|\Psi\> \propto \iint \ud \w\ud \w' \Phi(\w,\w') \;  |\w\>_\tx{s} |\w'\>_\tx{i}  = \iint \ud t\ud t' \Psi(t,t') \; 
	|t\>_\tx{s} |t'\>_\tx{i}. 
	\L{psi}	
	\end{align}
Here, $|\w\>_\tx{s,i}$ and $|t\>_\tx{s,i}$ denote single-photon states with well-defined angular frequency $\w$ and detection time instant $t$, respectively. 
The frequencies are counted relative to the corresponding cavity resonance frequencies $\w_\tx{s,i}^{(0)}$.
The functions $\Phi(\w,\w')$ ($\Psi(t,t')$) define joint spectral (temporal) probability amplitudes to detect signal and idler photons with frequencies $\w$ and $\w'$ (at time instants $t$ and $t'$), respectively.
The functions are related via a two-dimensional Fourier transformation.
The state describes a situation when a generated photon pair has left the optical cavity and has not been lost due  to the intracavity losses.
To increase the probability of such events outcoupling rates of signal and idler photons should exceed intracavity loss rates, i.e. the cavity should be overcoupled for signal and idler fields \cite{Heebner2004}.

The joint spectral amplitude of generated photon pairs $\Phi(\w,\w')$ is modeled using the following general expression, which is applicable for the description of the SPDC and SFWM processes \cite{Garay-Palmett2013, KHLuo2015, Vernon2017}:
	\begin{align}
	\begin{split}
	& \Phi(\w,\w') = \a_\tx{p}(\w + \w' - \D) F_\tx{s}(\w) F_\tx{i}(\w'), \L{Phi}
	\\
	& \text{where} \; F_\tx{s,i}(\w) = (\g_\tx{s,i}/2-i\w)^{-1} \, .
	\end{split}
	\end{align}
For SPDC, $\a_\text{p}(\w)$ describes the spectral amplitude of the field which pumps the nonlinear medium. Then photon generation rate depends linearly on the pump power. For SFWM, $\a_\text{p}(\w)$ should be replaced by the following convolution of the field amplitude: $\int \a_\tx{p}(\w') \a_\tx{p}(\w-\w') \ud\w'$, and the generation rate depends quadratically on the pump power.
The pump field is assumed to be strong enough to be treated as classical and small enough to neglect generation of multiple photon pairs within the time interval of interest.
The pump amplitude can be enhanced by the presence of a cavity for the pump field in a triply resonant configuration (see \ref{appx:triply} for the description of the enhancement).
The frequency mismatch $\D$, appearing in (\ref{Phi}), should be substituted by (\ref{d}) or (\ref{d1}).
$F_\tx{s,i}(\w)$ are Lorentzian functions which describe the cavity resonances that have single maxima at frequencies $\w_\tx{s,i}^{(0)}$ and full widths at half maximum (FWHM) $\g_\tx{s,i}$.
More complicated spectral shapes of resonances can be included into the model modifying $F_\tx{s,i}(\w)$. 
For example, in a microring cavity, a double-spiked spectrum can occur due to the coupling of counter-propagating waves induced by surface roughness \cite{Weiss1995, Little1997}.

The joint temporal amplitude $\Psi(t,t')$, corresponding to (\ref{Phi}), is obtained via a two-dimensional Fourier transformation of $\Phi(\w,\w')$ and reads
	\begin{align}
	\begin{split}
	& \Psi(t,t') = \int \ud t'' \; \a^{j=1,2}_\tx{p}(t'') e^{-i \D t''} \G_\tx{s}(t-t'') \G_\tx{i}(t'-t''), 	\L{Psi_t} \\
	& \text{where} \; \G_\text{s,i}(t) = e^{-\g_\tx{s,i} t/2} \theta(t) \, .
	\end{split}
	\end{align}
Here $\a_\tx{p}(t) =  \int \ud\w \a_\tx{p}(\w) e^{-i \w t}$ represents a slowly-varying temporal amplitude of the pump field, which should be raised to the power 1 to describe the SPDC process and to the power 2 for the description of the SFWM process.
$\G_\text{s,i}(t)$ represent the impulse responses of the cavity at the signal and idler frequencies with the response rates defined by $\g_\tx{s,i}$. 

In the presented model, the frequency mismatch $\D$ appears equivalently for SPDC and SFWM processes. 
In the following, we consider properties of the generated fields in continuous and pulsed excitation regimes.
These results are valid for both SPDC and SFWM.

%%%%%%%%%%%%%%%%%%%%%%%%%%%%%%%%%%%	
\subsection{Continuous pump regime}\L{sec:CW}

\begin{figure}[htb]
  \centering
  \includegraphics[scale=1.3]{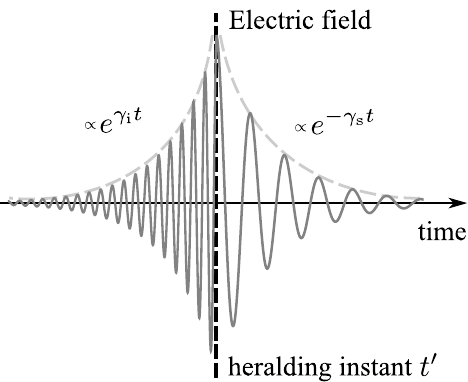} 
\caption{Temporal behavior of an electric field of a signal photon heralded from an entangled photon pair (\ref{Psi_tt}) upon detection of the idler photon at time instant $t'$. 
Time constant of the left rising exponent is $2\g_\tx{i}^{-1}$ and of the right decaying exponent is $2\g_\tx{s}^{-1}$ according to (\ref{Psi_tt}).
Frequency mismatch $\D$ leads to a phase modulation of the electric field such that the carrying frequency changes stepwise at the heralding instant by the value $\Delta$.
The same behavior holds for the electric field of idler photon, where $t$ and $t'$, as well as $\g_\tx{s}$ and $\g_\tx{i}$, should be exchanged.}
\label{fig:mode-t}
\end{figure}

In the continuous pump regime, the pump spectrum is monochromatic and the pump amplitude does not depend, in a good approximation, on time. One can put in the above expressions $\a_\tx{p}(\w) = \a_\tx{p} \dd(\w)$ and $\a_\tx{p}(t) = \a_\tx{p}$. Then the joint spectral amplitude of the generated signal-idler photon pairs reads:
	\begin{align}
	& \Phi(\w,\w') = \a_\tx{p}^{j=1,2} \dd(\w+\w'-\D) F_\tx{s}(\w) F_\tx{i}(\w').
	\end{align}
The frequencies of the photons are perfectly anti-correlated due to energy conservation.
The corresponding temporal joint amplitude of the photon pairs reads
	\begin{align}
	\Psi(t,t') = \frac{\a_\tx{p}^{j=1,2}}{\bar\g_\tx{si} - i\D} \times 
	\left\{\begin{array}{ll}
	e^{-\frac{\g_\tx{i}}{2}(t'-t)-i\D t}, & t<t'\\
	e^{-\frac{\g_\tx{s}}{2}(t-t')-i\D t'}, & t>t', 
	\end{array}\right.
	\L{Psi_tt}
	\end{align}
where $\bar\g_\tx{si} = (\g_\tx{s} + \g_\tx{i})/2$ is the average response rate of the cavity.
The joint amplitudes above can not be factorized with respect to their arguments  and the generated signal and idler photons are time-energy entangled \cite{MacLean2018}. 
The joint temporal amplitude can be interpreted as follows - it defines a temporal mode of signal photon heralded on the time-resolving detection of the idler photon at time instant $t'$, or, vice versa, it defines a temporal mode of the idler photon heralded upon the signal photon detection at time instant $t$ \cite{Averchenko2016}. 
A nonzero frequency mismatch $\D$ leads to the phase modulated temporal mode (i.e. time-dependent carrier frequency) of heralded photons that is shown schematically in Fig.~\ref{fig:mode-t}.
Thus the control over the frequency mismatch in the considered scheme is important to herald photons in a well-defined spectro-temporal mode.

We calculate the main characteristics of the generated signal and idler fields using the above expressions and consider the effect of the frequency mismatch on these characteristics.
The second-order temporal cross-correlation function of the generated fields can be calculated as \cite{Loudon2000}: 
	\begin{align}
	& g_\tx{si}^{(2)}(t,t') \propto \<\hat a_\tx{i}^{\+}(t') \hat a_\tx{s}^{\+}(t) \hat a_\tx{s}(t) \hat a_\tx{i}(t')\> \propto |\Psi(t,t')|^2.
	\L{Gsi-def}
	\end{align}
Using expression (\ref{Psi_tt}) one gets that the cross-correlation function depends on the time difference $\t = t'-t$:
	\begin{align}
	& g_\tx{si}^{(2)}(\t) \propto 
	\left\{\begin{array}{ll}
	e^{\g_\tx{s} \t}, & \t<0\\
	e^{-\g_\tx{i} \t}, & \t>0
	\end{array}\right. \, .
	\label{eq:Gsi}
	\end{align}
The cross-correlations are stationary for a continuous wave pump. They reduce exponentially with the time delay between the photocounts.
The temporal behavior is not affected by the frequency mismatch.
Measurement of the function (\ref{eq:Gsi}) yields the response rates $\g_\tx{s,i}$ of the signal and idler cavity modes.

The mean flux of signal photons is defined as: 
	\begin{align}
	& n_\tx{s}(t) \propto \<\hat a_\s^\+(t) \hat a_\tx{s}(t)\> \propto \int \ud t' |\Psi(t,t')|^2.
	\end{align}
The flux of idler photons is defined in a similar way, where the integration is performed over the first argument of the probability amplitude.
As a result, in the continuous pump regime the fluxes of signal and idler photons are constant and read:
	\begin{align}
	& n_\si \propto \frac{\text{P}_\tx{p}^{j=1,2}}{(\bar\g_\tx{si}^2 + \D^2)}.
	\L{n_CW}
	\end{align}
where P$_\tx{p} = |\a_\tx{p}|^2$ is the pump power.
The pump power is defined by the external pump in the doubly resonant configuration.
In a triply resonant configuration the power should be replaced with the average pump energy stored inside the pump mode of the cavity (see expression \ref{eq:n_p}).
Generated signal and idler fluxes depend linearly on the pump power for SPDC process and quadratically for SFWM process.
The fluxes have a Lorentzian dependence on the frequency mismatch $\D$ with the FWHM given by $2\bar\g_\tx{si}$, which is twice the average response rate of the cavity.

\begin{figure}[t]
  \centering
  \includegraphics[scale=1.2]{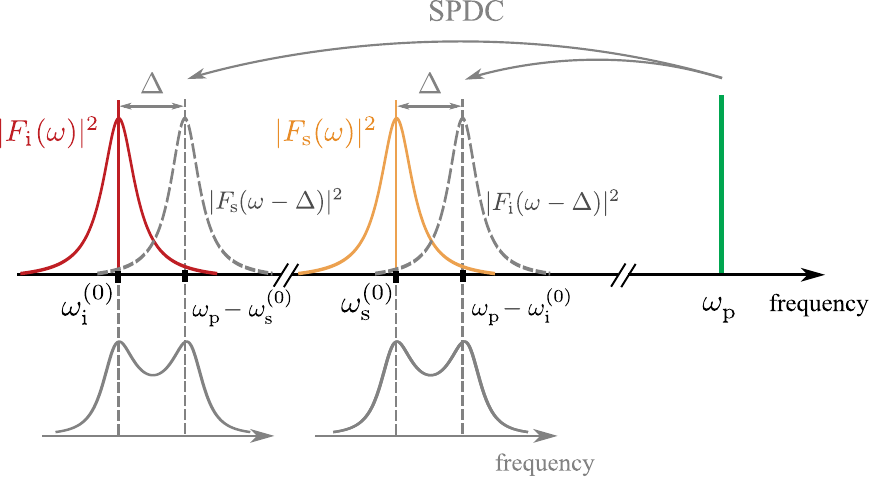} 
\caption{Formation of a double-peak spectrum of signal and idler fields in a cavity-assisted SPDC process at presence of a nonzero frequency mismatch $\Delta$. 
For the illustration purpose we assume equal linewidths for the signal and idler cavity resonances, i.e. $\g_\s \approx \g_\i$.
The parametric scattering of the monochromatic pump field on idler (quantum) vacuum mode with the Lorentzian spectrum (red curve) leads to the generation of signal field with the Lorentzian spectrum (grey dashed curve), which is shifted with respect to the signal resonance frequency by the frequency mismatch.
The overlap between two Lorentzian functions results in the double-spiked spectrum of the generated signal photons (see expression (\ref{S_si})).
The same mechanism leads to the double-peak spectrum of idler photons.
This mechanism also holds for SFWM.
}
\label{fig:mode-w}
\end{figure}

The spectral density of the generated signal photons is calculated as $S_\s(\w) \propto \<\hat a_\tx{s}^\+(\w) \hat a_\tx{s}(\w)\> \propto \int \ud \w' |\Phi(\w, \w')|^2$. The spectral density of idler photons is obtained similarly by integrating over the first argument of the joint spectral density. In the continuous pump regime, expressions for spectral densities of signal and idler photons generated in the parametric conversion read:
	\begin{align}
	& S_\tx{s,i}(\w) \propto |F_\tx{s,i}(\w)|^2 |F_\tx{i,s}(\w-\D)|^2.
	\L{S_si}
	\end{align}
Expressions (\ref{S_si}) describe spectra of single photon pulses generated both in the SPDC and SFWM process, where temporal shape of the pulses is shown in Fig.~\ref{fig:mode-t}.
Spectra are products of two Lorentzian functions shifted by $\D$ with respect to each other. For a frequency mismatch $\D =0$, the spectra $S_\tx{s,i}$ are squared Lorentzian functions and the FWHM is approximately equal to $0.64 \cdot \g_\tx{s,i}$ (if $\g_\tx{s} \approx \g_\tx{i}$) (see also \cite{KHLuo2015, Lu2016}). 
For a frequency mismatch $|\D| > \g_\tx{s,i}$, the spectra exhibit a double-peak structure. One peak resides at a cavity resonance and the other peak is shifted by the frequency mismatch $\Delta$.
Fig.~\ref{fig:mode-w} depicts the mechanism that leads to the formation of the double-peak spectrum of the signal photons in the SPDC process. 
The same mechanism leads to the double-peak spectrum of idler photons. This mechanism also holds in SFWM.
For the illustration we assumed that the cavity is symmetric.
One can show, that for an asymmetrical cavity, when $\g_\s \neq \g_\i$, the double-peak structure is asymmetric, namely, one of the peaks is broader than the other and has smaller height.
The double-spiked spectrum of the generated photons is a signature  of the phase-modulation of the single-photon wavepackets as seen in the expression (\ref{Psi_tt}) and depicted in Fig.~\ref{fig:mode-t}.
Notably, there is another, completely unrelated to the frequency mismatch $\Delta$, mechanism that can lead to a two-component structure of the photons spectrum in microring cavities, namely, splitted spectral profiles of the cavity resonances due to the backscattering on surface roughness  \cite{Weiss1995, Little1997}.
It is worth mentioning that in the work \cite{Drummond1990} an effect of the frequency mismatch on squeezing spectrum of the light produced from the nonedegenerate parametric oscillator below the oscillation threshold has been considered and double-peak structure of the spectrum has been predicted.

\begin{figure}
\center{\includegraphics[width=0.5\linewidth]{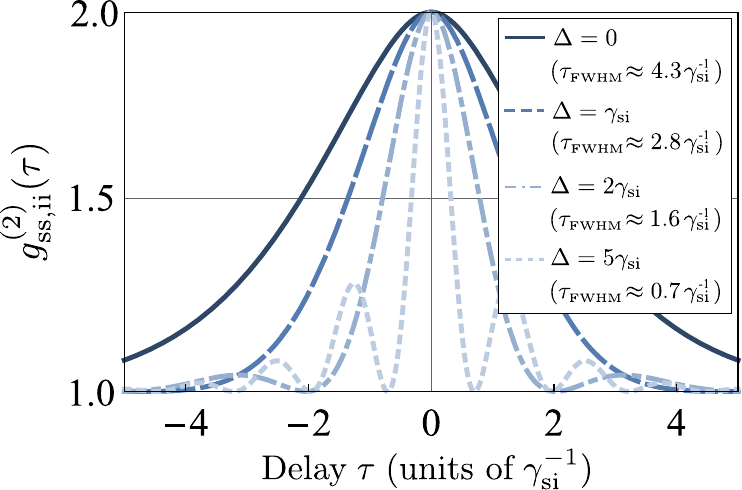}}
	\caption{Auto-correlation functions $g_\tx{ss,ii}(\t)$ of the signal (idler) field (see expression (\ref{eq:gssii})) generated in a cavity-assisted SPDC (or SFWM) under a continuous wave pump with different values of the frequency mismatch $\D$. For each curve we denote the FWHM of the auto-correlation time in units of $\g_\tx{si}^{-1}$.}
	\label{fig:gss}
	\end{figure}

The second-order auto-correlation function of signal (idler) field
characterizes the temporal distribution of photon pairs.
The function is nonzero if more than one photon pair is generated. 
Based on the Gaussian factorization theorem \cite{Bocquillon2009}, the normalized auto-correlation function for the signal field can be calculated as:
	\begin{align}
	\begin{split}
	& g_\text{ss}^{(2)}(t,t') = 1 + \frac{|\<\hat a_\tx{s}^\+(t) a_\tx{s}(t')\>|^2}{n_\s(t) n_\s(t')}, \L{Gss-def}\\
	& \text{where} \; \<\hat a_\tx{s}^\+(t) a_\tx{s}(t')\> \propto \int \Psi^*(t,t'') \Psi(t',t'') \ud t'' \, .
	\end{split}
	\end{align}
The auto-correlation function for the idler field is defined by a similar expression, where the integration is performed over the first argument of the probability amplitude.
In the continuous pump regime the auto-correlation functions for signal and idler fields depend on the time difference $\t = t'-t$. Particularly for $\g_\tx{s}= \g_\tx{i} = \g_\tx{si}$ the functions coincide and read (for the general case see expression \ref{g2_general})
	\begin{align}
	& g_\tx{ss,ii}^{(2)}(\t) = 1 + \(\cos\[\frac{\D \cdot \t}{2}\] + \frac{\g_\tx{si}}{\D} \sin\[\frac{\D \cdot |\t|}{2}\]\)^2 e^{-\g_\tx{si}|\t|} \, .	\label{eq:gssii}
	\end{align}
These correlation functions are stationary and do not depend on the pump intensity in the limit when there is less than photon pair in a cavity at any point of time.
Dependence of the correlation functions peaks (i.e., $g_\tx{ss,ii}^{(2)}$) on the pump power, omitting their temporal (frequency) properties, is derived, particularly, in the work \cite{McNeil1983}.
Fig.~\ref{fig:gss} shows the auto-correlation function for different values of the frequency mismatch $\D$.
Since $g_\tx{ss,ii}^{(2)}(0) > g_\tx{ss,ii}^{(2)}(\t \neq 0)$, the functions reveal super-Poissonian statistics and bunching of the generated photon pairs, which is typical for photon statistics of chaotic light \cite{Loudon2000}.
In the limit $\D \rightarrow 0$, the auto-correlation functions take the form $g_\text{ss,ii}^{(2)}(\t) = 1 + \(1 +\g_\tx{si} |\t|/2\)^2 e^{-\g_\tx{si}|\t|}$, analogous to a result of \cite{Luo2015, Lu2016}. The functions are not exponential and the auto-correlation time measured at FWHM is longer than the cross-correlation time of the fields defined as $\g_\tx{si}^{-1}$. 
In the case $\D\neq0$, the auto-correlation FWHM decreases, and oscillations of the functions with the period $2\pi/\D$ appear. 
The auto-correlation functions and the corresponding auto-correlation times at FWHM level are shown in Fig.~\ref{fig:gss} for different values of the frequency mismatch. 
The oscillations of the functions can be explained by the periodic transition between the parametric amplification and deamplification in the presence of the frequency mismatch.
The value of the auto-correlation time $\t_\text{FWHM}$ (expressed in units $\g_\text{si}^{-1}$) can be used to estimate the frequency mismatch. 
For $\Delta = 0$, the value should be approximately 4.3 (c.f. Fig. 3).
We analyzed several experiments \cite{Scholz2009, Fortsch2013, Fortsch2014, KHLuo2015} and found that the ratio varies from $1.6$ to $3.6$ which can be a signature of non-zero frequency mismatch in these experiments
\footnote{
We should also acknowledge that the spectral linewidth and correlation time are affected by parametric gain, which depends on the pump power. Particularly, the linewidth is becoming nearly monochromatic as the threshold of parametric oscillations is approached. So comparing ratios from different experiments requires a knowledge of the parametric gain and without it may be not very reliable.}.

%%%%%%%%%%%%%%%%%%%%%%%%%%%%%%%%%%%
\subsection{Pulsed pump regime}\L{sec:pulsed}

Here we consider excitation of cavity-assisted parametric conversion with pump pulses of arbitrary temporal profile  in the framework of  the slow varying amplitude approximation which is most commonly used in coupled modes analysis. This approximation is discussed in \ref{appx:triply}. A more accurate description of the resonator's response to a rapidly changing pump field would require using time-delayed equations rather than differential equations such as (\ref{B1}) and (\ref{eq:rateequationsshort}) (see, for example, \cite{Averchenko2011}).

For doubly resonant configuration the pump amplitude is defined by the pump laser. For triply resonant configuration the pump amplitude is modified by the pump cavity (see \ref{appx:triply}).
The fluxes of the generated signal and idler photons depend on time and the corresponding general expressions can be obtained from Eq. (\ref{Psi_t}):
	\begin{align}
	& n_\tx{s,i}(t) \propto 
	\iint\limits \a_\tx{p}^{*j}(t') \a_p^{j}(t'') e^{i \D (t'-t'')} \G_\tx{s,i}(t-t') \G_\tx{s,i}(t-t'')  e^{-\frac{\g_\tx{i,s}}{2} |t'-t''|} \ud t' \ud t'', \L{nsi_t}
	\end{align} 
where we remind that 
$j=1$ stands for SPDC process and $j=2$ for SFWM process.
The expressions are applicable for arbitrary temporal shapes of pump pulses.

\begin{figure}
\center{\includegraphics[width=0.5\linewidth]{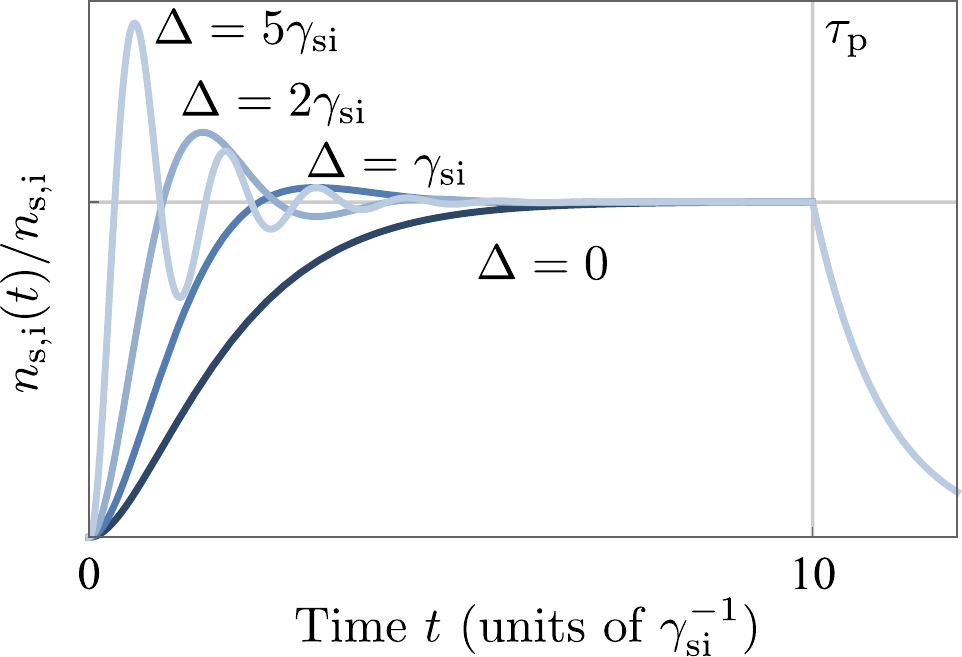}}
	\caption{Normalized time-dependent flux $n_\text{s,i}(t)$ of signal/idler photons generated under the excitation of the cavity-assisted SPDC or SFWM with a rectangular pump pulse of the duration $\t_\text{p} = 10 \g^{-1}_\tx{si}$. The plots are normalized to the stationary flux $n_\text{s,i}$ and are plotted for different frequency mismatches $\D$.}
	\label{fig:ns}
	\end{figure}

As an example, we consider rectangular pulses of duration $\t_\tx{p}$ which pump a nonlinear medium. We also assume that $\g_\tx{s}=\g_\tx{i}=\g_\tx{si}$. Then time-dependent fluxes of signal and idler photons read
	\begin{align}
	\begin{split}
	& n_\tx{s,i}(t) = n_\si \times \left\{\begin{array}{ll}
	\varphi(t), & 0 \leq t \leq \t_\tx{p}\\
	\varphi(\t_\tx{p}) e^{-\g_\tx{si}(t-\t_p)}, & \t_\tx{p} < t 
	\end{array}\right. \, ,\\
	& \text{where}\, \varphi(t) = 1-e^{-\g_\tx{si} t}\(\cos(\D \cdot t) + (\g_\tx{si}/\D) \sin(\D \cdot t)\).
	\label{eq:nst}
	\end{split}
	\end{align}
Here $n_\si$ are steady-state photon fluxes generated in the continuous pump regime (\ref{n_CW}). The function $\varphi(t)$ describes the temporal behavior of the photon fluxes and it is shown in Fig.~\ref{fig:ns}. 
Increasing frequency mismatch leads to faster rise times of the photon fluxes, as well as, to oscillations with the angular frequency equal to $\D$. 
At the end of the pump pulse the photon fluxes undergo the free cavity ringdown with ringdown times defined by the inverse response rates $\g_\tx{s,i}^{-1}$ of the cavity.
This dynamics of parametric excitation of the signal/idler cavity mode is qualitatively similar to the dynamics of the driven excitation of a cavity mode, presented in \ref{appx:triply} (see  expression (\ref{eq:npt}) and Fig.\ref{fig:pump}).

Notably, when the pump pulses are shorter than the response times of the cavity at signal and idler resonance frequencies (this requirement can be also fulfilled in a triply resonant configuration \cite{Vernon2017} for a high response rate of the pump cavity resonance) the generated photons are not entangled. Indeed, the joint temporal amplitude (\ref{Psi_t}) factorizes with respect to $t$ and $t'$, when ${\cal \a}_\text{p}(t) \rightarrow  \delta(t)$. Non-entangled photon pairs can be used for the efficient heralding of single-mode photons \cite{Mosley2008} with the nanosecond coherence times and  exponentially decaying temporal shape.

Second-order cross-correlation $g_\tx{si}^{(2)}(t,t')$ and auto-correlation $g_\tx{ss,ii}^{(2)}(t,t')$ functions of fields generated in the pulsed regime are non-stationary and depend on both time instants $t$ and $t'$. Calculation of the functions for particular pump pulses can be done using the expressions (\ref{Gsi-def}) and (\ref{Gss-def}), respectively.

%\FloatBarrier
%%%%%%%%%%%%%%%%%%%%%%%%%%%%%%%%%%%
\section{Experimental scheme}
\begin{figure}[htb]
  \centering
  \includegraphics[scale=1.0]{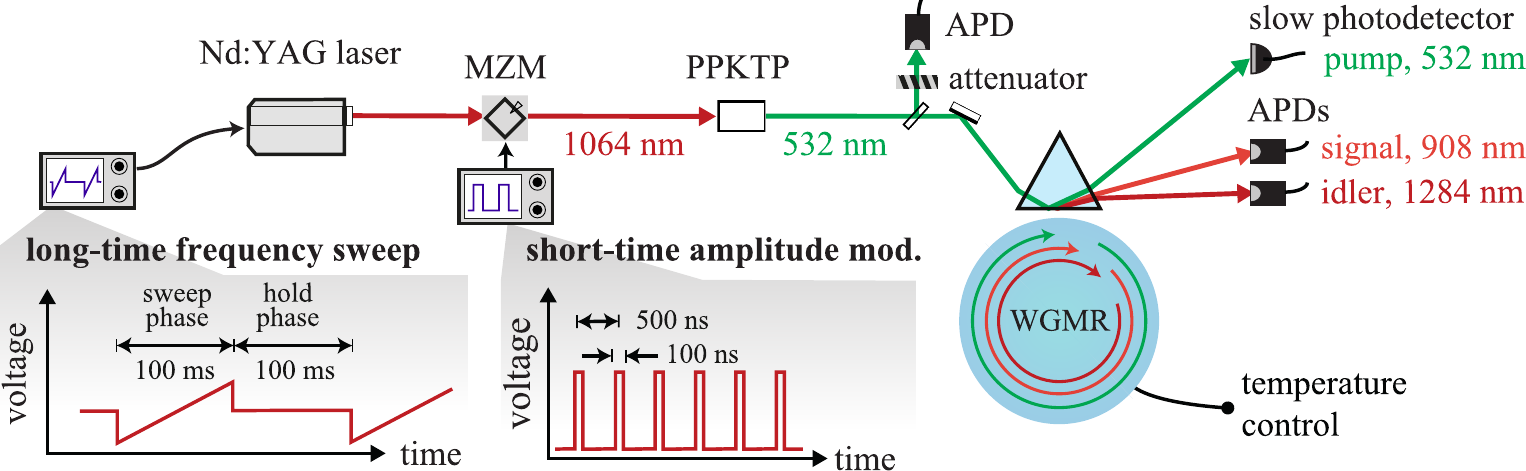} 
\caption{Experimental setup for cavity-assisted parametric down-conversion with pulsed excitation. We generate photon pairs via parametric down-conversion in a whispering-gallery mode resonator (WGMR). For generating the resonator pump light at a wavelength of \SI{532}{\nm}, we frequency-double a Nd:YAG laser (wavelength at \SI{1064}{\nm}) in a periodically poled potassium titanyl phosphate crystal (PPKTP). We carve out nanosecond pump pulses at the subharmonic frequency by means of a Mach-Zehnder modulator (MZM). The attenuated pump pulses are monitored at the harmonic frequency with a Si avalanche photodetector (APD) before the resonator (see pump pulses in Fig.~\ref{fig:pulsedcountsnano}). The signal and idler photons are detected with a Si APD and a InGaAs/InP APD, respectively. At the millisecond scale, we use a slow silicon photodetector to monitor the pump light in reflection from the resonator (see Fig.~\ref{fig:bandwidthmeas}(a)) and employ a sweep-and-hold lock for the pump laser frequency.
}
\label{fig:pulsingsetup}
\end{figure}

\L{sec:scheme}
In this section, we describe our experimental setup for parametric down-conversion in a nonlinear whispering gallery mode resonator (WGMR) driven by a pulsed pump. 
The basic experimental setup is shown in Fig.~\ref{fig:pulsingsetup} and described also in Ref.~\cite{Furst2010natural,Furst2010,Michael2013,Schunk2015a,Schunk2016}. Our WGMR is made of MgO-doped ($5.8 \%$) lithium niobate with a major radius $R$ and rim curvature $\rho$ of \SI{2.5}{\mm} and \SI{0.58}{\mm}, respectively (see Ref.~\cite{Furst2010natural,Schunk2016} for more details on phase-matching in the WGMR). We use a frequency-doubled Nd:YAG laser at a wavelength of \SI{532}{\nm} to pump the PDC process. We couple pump light to the WGMR via frustrated total internal reflection in a diamond prism. The same prism enables out-coupling of the parametric light. More advanced techniques allow for an individual coupling of the pump and the parametric light via polarization-selective out-coupling \cite{SedlmeirSelective2016}.

We generate square pulses (pulse duration of \SI{100}{\ns}, pulse period of \SI{500}{\ns}) from a Nd:YAG laser at a wavelength of \SI{1064}{\nm} by means of an amplitude modulator, i.e. a Mach-Zehnder modulator (Modulator 10020490-901 from JDSU, bandwidth of \SI{3}{\GHz}) driven by an arbitrary waveform generator (AWG7061B from Tektronix, bandwidth of \SI{6}{\GHz}).
The pulses are frequency-doubled in a periodically poled potassium titanyl phosphate (PPKTP) crystal to a wavelength of \SI{532}{\nm} and measured with a Si avalanche photodetector (APD) (SPCM-NIR from Perkin Ellmer). 
The intensity at the harmonic frequency depends quadratically on the intensity at the subharmonic frequency. A modulation of the light already at the subharmonic frequency increases the peak to valley ratio and shortens the rise time of the pulses.
 
We sweep the central frequency of the frequency-doubled Nd:YAG laser during \SI{100}{\ms} across \SI{105}{\MHz}. The sweep phase is followed by a \SI{100}{\ms}-hold phase with a constant pump laser frequency locked to the pump mode resonance frequency.

\begin{figure}[htb]
  \centering
  \includegraphics[scale=1.0]{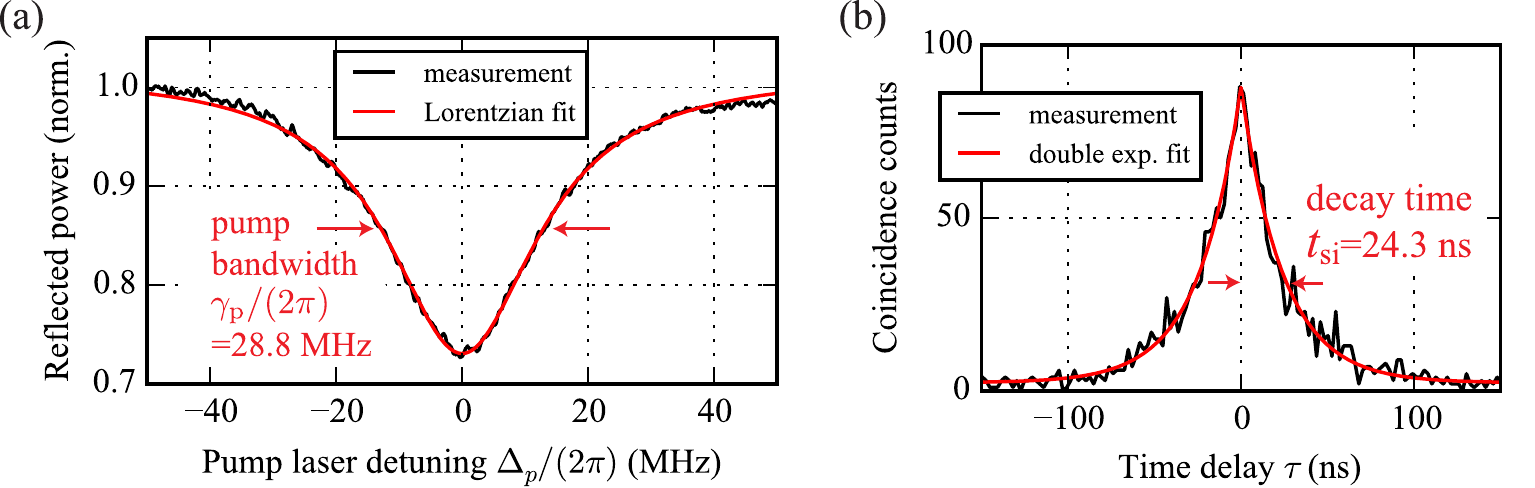} 
\caption{Bandwidth measurements of the pump and the parametric modes. (a) The Lorentzian fit of the reflected pump light (see Eq.~\ref{eq:Pout}) measured with a slow photodetector gives a bandwidth of the pump resonance $\gamma_\tx{p} / (2\pi)= \SI{28.8}{\MHz}$ and the pump decay time $t_\tx{p}=1 / \gamma_\tx{p}  \approx \SI{5.5}{\ns}$. (b) The symmetric double exponential fit $\propto \tx{exp} (-|\tau|/t_\tx{si})$ (see Eq.~\ref{eq:Gsi} for the cross-correlation function) of the measured signal and idler coincidence histogram (measurement time of \SI{50}{\s}, binning  time of \SI{2.2}{\ns}) yields the signal-idler decay time $t_\tx{si}=\gamma_\tx{si}^{-1}=\SI{24.3}{\ns}$ and the bandwidth $\gamma_\tx{si}/(2\pi)  \approx \SI{6.5}{\MHz}$. Here we assume equal bandwidths for signal and idler resonances.}
\label{fig:bandwidthmeas}
\end{figure}

We use a slow silicon photodetector (PDA36A-EC from Thorlabs, bandwidth of \SI{45}{\kHz}) to monitor the pump light in reflection from the resonator. A Lorentzian fit (see Eq.~\ref{eq:Pout}) of the reflected pump light during the sweep phase (see Fig.~\ref{fig:bandwidthmeas}a) yields a resonator bandwidth of $\gamma_\tx{p} / (2\pi)=\SI{28.8}{\MHz}$ measured at critical coupling for the pump \cite{Schunk2016}.
  
In order to predict the conversion channels of PDC \cite{Breunig2016,Strekalov2016jop}, we initially perform an analysis of the eigenmodes of the WGMR at the pump wavelength \cite{Schunk2014a}. For a maximal conversion efficiency \cite{Furst2010} and single mode operation \cite{Michael2014} at the single photon level, we pump the WGMR and generate parametric light in fundamental modes \cite{Breunig2013b,Schunk2014a}. The resonator temperature is stabilized at the millikelvin scale around a temperature of \SI{133}{\degreeCelsius}. The signal and idler wavelengths are \SI{908}{\nm} and \SI{1284}{\nm}, respectively. A pump threshold of $\tx{P}_\tx{th}=\SI{18}{\uW}$ was measured for a signal (idler) wavelength of \SI{895}{\nm} (\SI{1312}{\nm}) \cite{Schunk2016} in CW pump regime. 

For an operation below threshold, we pump the resonator with approximately $\tx{P}^\tx{in}_\tx{p} = \SI{0.28}{\uW}$ average coupled power measured with a low-bandwidth powermeter. The power of each pulse is 5 times higher according to the adjusted ratio of pulse period and pulse duration. We use Si APDs (SPCM-NIR from Perkin Ellmer) for detection of the attenuated pump pulses and the signal photons. For detection of the idler, we use an InGaAs/InP avalanche photodetector (ID220 from ID Quantique). In Fig.~\ref{fig:bandwidthmeas}(b), we show the signal and idler coincidences measured for \SI{50}{\s} selecting only photons during the hold phase of the pump. Here, we fit the coincidences with a double exponential function according to Eq.~\ref{eq:Gsi} under the assumption of equal bandwidths of the signal and idler resonances. The signal-idler decay time $t_\tx{si}=\SI{24.3}{\ns}$ is 4.4 times longer than the pump decay time $t_\tx{p}=\SI{5.5}{\ns}$. 
These parameters allow us to modulate the intracavity pump field faster than the response time of the parametric process which is defined by $t_\tx{si}$.
Hence, we can freely control the dynamics of the parametric process up to the limits of the fastest external modulation frequency.

\begin{figure}%[htb]
  \centering
  \includegraphics[scale=1.0]{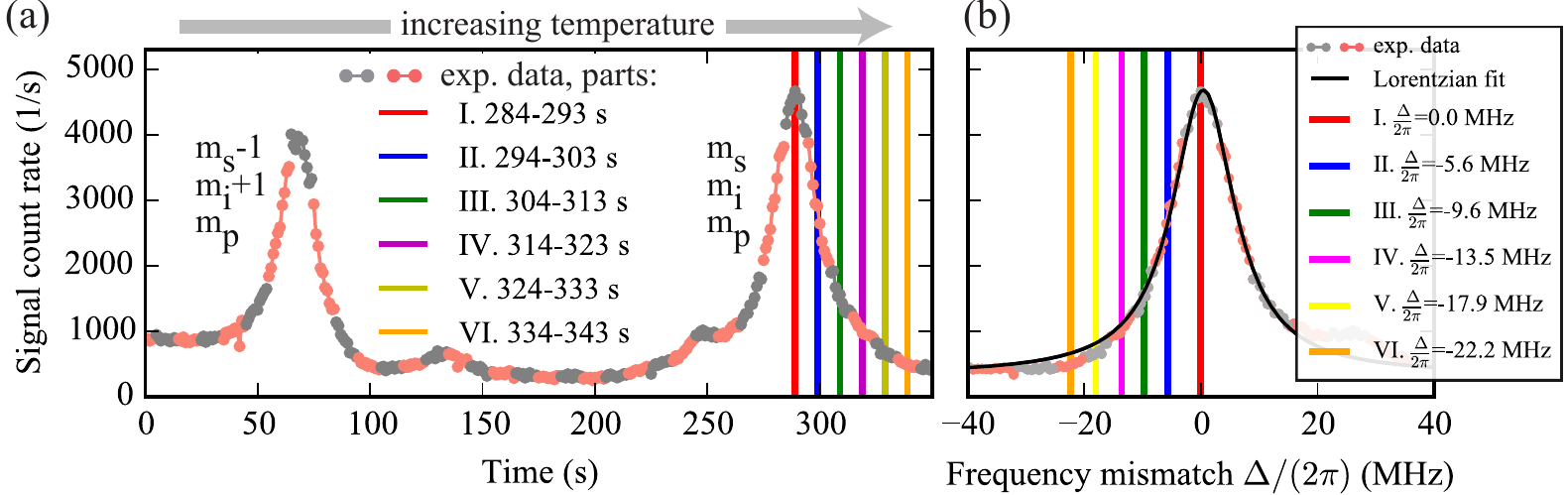} 
\caption{
(a) Temperature-dependent phase matching of parametric down-conversion below threshold. We continuously increase the resonator temperature while recording the signal photons during the hold phase of the laser frequency sweep. 
We divide the measured trace in 10 s segments (alternating grey-pink colors). 
Along each segment the resonator properties, such as temperature, are considered to be constant.
(b) The increase in the resonator temperature corresponds to a decrease in the resonance frequencies $\w_\tx{p,s,i}^{(0)}$ and to a decrease in the frequency mismatch of parametric down-conversion $\Delta$, which passes through zero and becomes negative. We calibrate the x-axis in terms of the frequency mismatch $\Delta$ by fitting the second peak in (a) with a Lorentzian function  according to Eq.~\ref{n_CW} and setting its FWHM to $2\gamma_\tx{si}/(2\pi)=\SI{13.1}{\MHz}$ (see  Fig.~\ref{fig:bandwidthmeas} (b)).
We also select six consecutive 10-s segments and tag their centers with the corresponding estimated values of the frequency mismatch (see vertical colored lines).
For each tagged segment we perform the pump frequency sweep measurement and pulsed pump measurement, whose results are presented in Fig.~\ref{fig:lockingsignal} and Fig.~\ref{fig:pulsedcountsnano}, respectively.
}
\label{fig:tempcounts}
\end{figure}

For studying the impact of the frequency mismatch $\Delta$ on the generation rate of photon pairs, we measure the signal photon count rate while linearly increasing the resonator temperature by approximately \SI{20}{\milli\kelvin} (see Fig.~\ref{fig:tempcounts}). During the scan the laser central frequency is locked to the pump mode resonance frequency. 
The two pronounced peaks in the count rate represent a conversion to adjacent longitudinal modes for signal and idler \cite{Schunk2015a,Schunk2016}. The longitudinal modes of pump, signal, and idler are labeled by their respective azimuthal mode number $\tx{m}_\tx{p,s,i}$. Different smaller peaks in the signal photon count rate indicate the presence of conversion channels to higher-order whispering gallery modes. The pump laser absolute frequency follows the temperature-induced frequency shifts of the pump mode. The frequency mismatch $\Delta$ is proportional to the resonator temperature due to the linear thermo-refraction \cite{Schlarb1994} and the linear thermal expansion \cite{Weis1985} of the resonator
\footnote{
The theoretical proportionality factors for frequency tuning by temperature are approximately \SI{-27}{\MHz\per\milli\kelvin} for the pump mode and \SI{-6.3}{\MHz\per\milli\kelvin} for the parametric modes~\cite{Schlarb1994,Weis1985}. Hence, the frequency mismatch $\Delta/(2\pi)$ is changed by \SI{-14.4}{\MHz\per\milli\kelvin} for $\Delta_\tx{p}=0$}.
Hence, there is a proportionality factor between the measurement time and the frequency mismatch $\Delta$. 
We fit the second peak in Fig.~\ref{fig:tempcounts}(a) with the Lorentzian function whose bandwidth is twice the signal-idler bandwidth, i.e. $2\gamma_\tx{si} / (2\pi)=\SI{13.1}{\MHz}$ according to Eq.~\ref{n_CW}. 
The results in Fig.~\ref{fig:tempcounts}(b) show the signal count rate depending on the frequency mismatch $\Delta$. The respective frequency mismatches $\Delta / (2\pi)$ for the six parts range from \SI{0}{\MHz} (part I) to \SI{-22.2}{\MHz} (part VI). The graph is mirrored compared to Fig.~\ref{fig:tempcounts}(a) due to the negative dependence of the frequency mismatch $\Delta$ on the resonator temperature.

\FloatBarrier

%%%%%%%%%%%%%%%%%%%%%%%%%%%%%%%%%%%
\section{Determination of the frequency mismatch $\Delta$ via the pump frequency sweep}\L{sec:delta}

In the previous section the signal and idler measurements have been performed under the condition that the central frequency of pump laser $\w_\tx{p}$ is locked to the pump mode resonance frequency $\w_\tx{p}^{(0)}$ of the WGMR.
Then the frequency mismatch defined as $\D = \w_\tx{p} - \w_\tx{s}^{(0)} - \w_\tx{i}^{(0)}$ coincides with the frequency mismatch between the cavity resonance frequencies defined as $\D^{(0)} = \w_\tx{p}^{(0)} - \w_\tx{s}^{(0)} - \w_\tx{i}^{(0)}$.
In this section we sweep the pump central frequency while keeping the temperature and therefore the frequency mismatch $\D^{(0)}$ constant corresponding to one of the six values deduced in the previous section (see Fig.~\ref{fig:tempcounts}(b)).

Expressions for the photon count rates measured in our cavity-assisted triply resonant SPDC can be obtained from (\ref{n_CW}) replacing pump power with the intracavity pump energy (\ref{eq:n_p}). 
They read
	\begin{align}
	& n_\tx{s,i}(t) \propto \frac{1}{\bar\g_\tx{p}^2/4 + \D_\tx{p}^2} \;  \frac{1}{\g_\tx{si}^2 + \D^2},
	\L{nsi_WGMR}
	\end{align} 
where $\D_\tx{p} = \w_\tx{p}- \w_\tx{p}^{(0)}$ is a detuning of the pump field frequency from the cavity resonance frequency, and we assume $\g_\tx{s} = \g_\tx{i} = \g_\tx{si}$.
The first Lorentzian function describes the frequency dependent enhancement of the pump field in the WGMR.
The second Lorentzian function describes the efficiency of the parametric process depending on the mismatch from the parametric resonance. 
The efficiency is maximal, when the pump field is resonant with the parametric modes $\D=0$. 
On the other hand, since $\D = \D^{(0)} + \D_\tx{p}$, the efficiency is maximal when the condition $\D_\tx{p} = - \D^{(0)}$ is met during the sweep phase of the pump frequency.
This condition allows us to determine the frequency mismatch $\Delta^{(0)}$ by measuring the photon count rates depending on the pump laser frequency, and normalizing them to the intracavity pump power.

\begin{figure}[htb]
  \centering
  \includegraphics[scale=0.9]{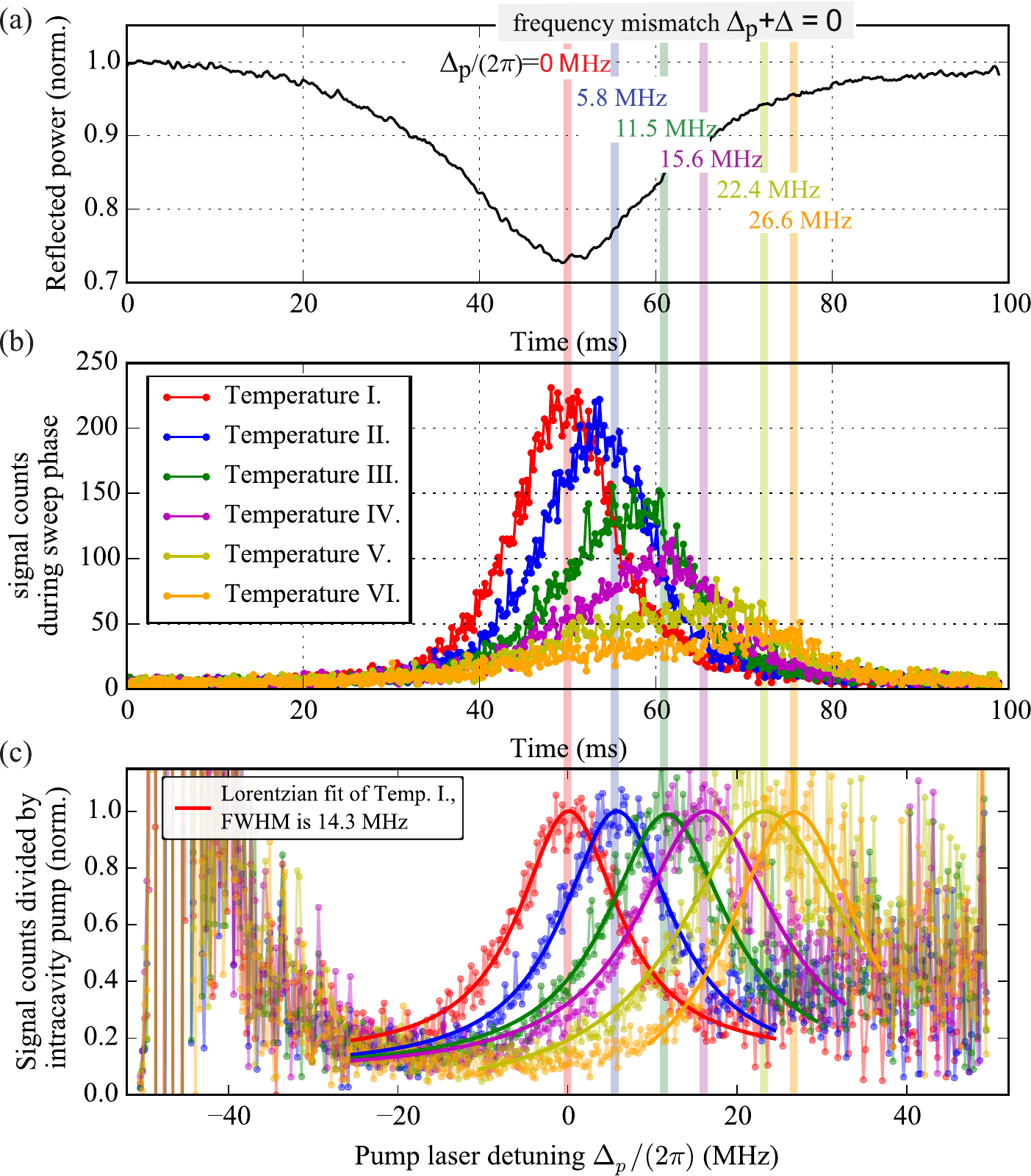} 
\caption{
Estimation of the frequency mismatch via the pump frequency sweep.
(a) We sweep the pump laser frequency for \SI{100}{\ms} across $100~\tx{MHz}$ symmetrically around the pump mode (see the sweep phase in Fig.~\ref{fig:pulsingsetup}) and measure the reflected power $P^\tx{out}_\tx{p}$ (see Eq.~\ref{eq:Pout}). (b) The maximum of the count rates shifts to a higher pump laser detuning $\Delta_\tx{p}$ for an increasing resonator temperature from part I to VI (see Fig.~\ref{fig:tempcounts}). (c) The signal count rates normalized to the intracavity pump energy E$_\tx{p}$ (see Eq.~\ref{eq:n_p}) show the detuning $\Delta_\tx{p}$ of the phase-matched ($\Delta=0$) pump laser frequency. The vertical bars show the central frequency of the respective Lorentzian (see Eq.~\ref{n_CW}), i.e. the frequency mismatch $\Delta$ during the hold phase of the pump laser ($\Delta_\tx{p}$=0).}
	\label{fig:lockingsignal}
\end{figure}

In Fig.~\ref{fig:lockingsignal}(a), the pump laser frequency sweep is shown as a function of time (see Fig.~\ref{fig:bandwidthmeas}(a) for a calibrated frequency axis).
The signal photon counts in Fig.~\ref{fig:lockingsignal}(b) are obtained during the sweep phase of the pump laser (see Fig.~\ref{fig:pulsingsetup} for the experimental scheme).
Six curves of the signal counts correspond in good approximation to different fixed resonator temperatures marked I - VI in Fig.~\ref{fig:tempcounts}, and hence to fixed frequency mismatches $\Delta^{(0)}$.
Curve I is symmetric around the pump resonance frequency $\w _\tx{p}$, yielding the highest count rates. 
An ideal frequency matching $\Delta =0 $ in this case is achieved for a zero pump laser detuning. 
Hence, curve I was measured exactly at the phase matching temperature, where the resonance frequencies of the mode triplet fulfill energy conservation $\w_\tx{p}^{(0)} = \w_\tx{s}^{(0)} + \w_\tx{i}^{(0)}$.

For a higher resonator temperature, the pump and parametric resonance frequencies shift to lower values at different rates, leading to a nonzero frequency mismatch.
Maximal count rates are now achieved when pump frequency is detuned from the cavity resonance. 
The intracavity pump power, however, is reduced for nonzero pump laser detuning, which in turn lowers the pair generation rate according to the $1/\left(\g^2_\tx{p}/4 + \Delta^2_\tx{p} \right)$ in Eq.~\ref{nsi_WGMR}.
In Fig.~\ref{fig:lockingsignal}(c), we take this factor into account and, following Eq.~\ref{eq:Pout}, we divide the signal count rates from Fig.~\ref{fig:lockingsignal}(b)) by the normalized intracavity pump power 1 - $P^\tx{out}_\tx{p}/P^\tx{in}_\tx{p}$  from  Fig.~\ref{fig:lockingsignal}(a).
 The resulting graph shows the pump frequency-dependent phase matching of PDC at different resonator temperatures. 
The maximum of the Lorentzian fit (see Eq.~\ref{n_CW}) of each trace allows us to deduce the frequency mismatch between the cavity resonance frequencies using the equality $\D^{(0)} = - \D_\tx{p}$.
There is a difference between these values and the ones deduced from the Lorentz approximation of the pump resonance in Fig.~\ref{fig:tempcounts}(b).
The difference is from 3 to 20 percent.
It can be due to the calibration innaccuracy of the horizontal axis in Fig.~\ref{fig:tempcounts}(b) and reduced signal-to-noise ratio of measurements performed at higher parametric resonance  mismatches (see Fig.~\ref{fig:lockingsignal}(c)).

According to Eq.~\ref{nsi_WGMR}, the FWHM of the normalized signal counts is equal to twice the signal-idler bandwidth $2 \gamma_\tx{si}$  The coincidence measurement (see Fig.~\ref{fig:bandwidthmeas}(b)) of signal and idler gives an expected FWHM of $2 \gamma_\tx{si} / (2\pi) = \SI{13.1}{\MHz}$, which is close to the fitted FWHM of \SI{14.3}{\MHz} (see curve I in Fig.~\ref{fig:lockingsignal}(c)).

\FloatBarrier
%%%%%%%%%%%%%%%%%%%%%%%%%%%%%%%%%%%
\section{Impact of the parametric resonance mismatch $\Delta$ on the parametric photons' rise time in pulsed excitation}\L{sec:exp_pulsed}
\label{sec:pulses}
In this section, we investigate the dynamics of the cavity-assisted SPDC below the oscillation threshold using our knowledge on the modes' bandwidths and the frequency mismatch gained in the previous sections. 
We study the dynamics of the signal photon count rate relative to the pump pulses for different temperature-controlled values of the parametric resonance frequency mismatch $\Delta$.

\begin{figure}[htb]
	\centering
	\includegraphics[scale=0.95]{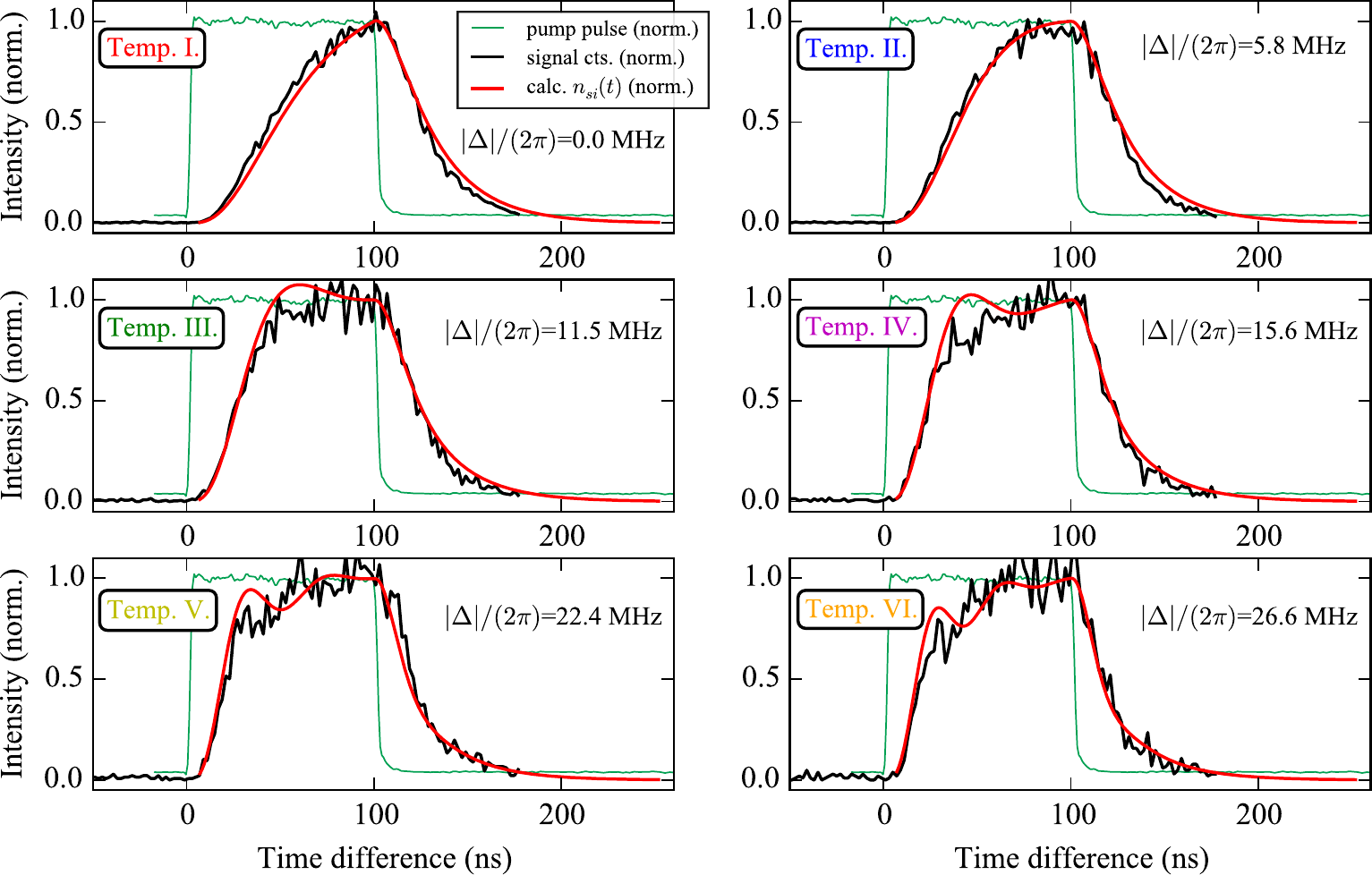} 
	\caption{Signal photon count rates with pulsed excitation. We send square pump pulses (green traces) with a pulse duration of \SI{100}{\ns} to the resonator to drive parametric down-conversion below the threshold (see Fig.~\ref{fig:pulsingsetup} for the experimental setup). The signal counts are evaluated during the hold phase of the pump laser frequency sweep (binning time of \SI{2}{\ns} of the histograms). The laser frequency is locked to the pump mode resonance frequency ($\Delta_\tx{p}=0$). The subfigures correspond to the six evaluation parts from Fig.~\ref{fig:tempcounts}(b), which are recorded for different temperature-induced frequency mismatches $\Delta$. At the beginning of the pulses, we observe shorter rise times of the signal photon count rate for an increased absolute frequency mismatch $|\Delta|$. The calculated cavity ring-down of the signal photons at the end of the pulses (signal-idler decay time is ${t_\tx{si}=\SI{24.3}{\ns}}$, see Fig.~\ref{fig:bandwidthmeas}) agrees well with the measured data.
}
	\label{fig:pulsedcountsnano}
\end{figure}

For Fig.~\ref{fig:pulsedcountsnano}, we pump the resonator with a squared pulse (green traces) and measure signal photon counts relative to the pump pulse
for six consecuteve 10-s segments of the cavity temperature scan (see vertical colored lines in Fig.~\ref{fig:tempcounts}(a)).
The signal counts are recorded during the hold phase of the pump laser, i.e. when the pump is locked on the resonance.
We also calculate and plot the normalized count rate profile from Eq.~\ref{eq:nst}  based on the measured bandwidths (see Fig.~\ref{fig:bandwidthmeas}) and the absolute frequency mismatch $|\Delta|$ known from the temperature sweep measurement in Fig.~\ref{fig:tempcounts}(b). The count rate profile is insensitive to the sign of the frequency mismatch $\Delta$.

At the beginning of the pump pulse, photon pairs are generated from an exponential loading ($n_\tx{p} \propto \left( 1 - e^{-t / t_\tx{p}}\right)$) of the pump field according to Eq.~\ref{eq:npt}. The normalized signal count histograms in Fig.~\ref{fig:pulsedcountsnano} also exhibit an exponential loading curve, but with a rise time strongly influenced by the frequency mismatch $\Delta$, which is consisted with Fig.~\ref{fig:ns}. The highest photon rates with the longest rise times are achieved in part I of Fig.~\ref{fig:pulsedcountsnano} for a zero frequency mismatch. At a nonzero frequency mismatch, the photon rates and rise times are decreased. We attribute this effect to increase of the parametric bandwidth with the absolute frequency mismatch $|\Delta|$ (see Fig.~\ref{fig:mode-w}).

At the end of the pulse, the pump light couples out from the resonator with a decay time faster than the signal-idler decay time (see Fig.~\ref{fig:bandwidthmeas} for a measurements of the decay times). In the limit of no pump light in the resonator, the signal photons' decay is solely determined by the bandwidth of the signal mode. Our calculation with the signal-idler decay time of \SI{24.3}{\ns} reproduces in good approximation the measured cavity ring-down of the signal photons.

For an increased frequency mismatch $\Delta$, the dynamics of the calculated signal photon counts exhibits an oscillatory behavior with an increasing oscillation frequency. This oscillation is caused by an alternating excitation and de-excitation of the parametric intracavity light. 
The measured photon counts in Fig.~\ref{fig:pulsedcountsnano} show a discrepancy with our calculations for an increasing absolute mismatch $\Delta$. 
Several reasons can be assumed.
First, signal photon counts are evaluated during \SI{10}{\s} and in each evaluation part the frequency mismatch $\Delta$ comprises a range of values.
Simulations show that averaging over frequency mismatches causes oscillations to disappear, in general. 
However, averaging over several megahertz, that corresponds to the experiment, can not explain the absence of oscillations in the measured signal photon counts.
Second, we assumed above that bandwidths of signal and idler resonances coincide (see Fig.~\ref{fig:bandwidthmeas}b and caption to it).
One can show, using expressions (\ref{nsi_t}), that the discrepancy of bandwidths leads to a decrease in the amplitude of the oscillations and their disappearance.
Finally, theoretical results are obtained in the approximation described in \ref{appx:triply}, which requires revision when the system is excited by square pulses.

Notably, the measurements shown in Fig.~\ref{fig:pulsedcountsnano} must not be confused with the temporal mode of the single photons, since we perform an ensemble measurement over different temporal modes.
Time or frequency resolving detection of idler photons is required to herald signal photons in a well-defined spectro-temporal mode \cite{Du2015, Averchenko2016, Sych2015}.

\FloatBarrier
%%%%%%%%%%%%%%%%%%%%%%%%%%%%%%%%%%%
\section{Conclusion}\L{concl}

In this work we study spectro-temporal properties of narrowband photon pairs generated in a cavity-assisted spontaneous parametric down-conversion (SPDC) and spontaneous four-wave mixing (SFWM) in the presence of the frequency mismatch of the generated photons from the cavity resonances.
First, we use a simple, yet generic, model that describes in one formalism the cavity-assisted SPDC and SFWM for both the continuous and pulsed regime.
We show that the aforementioned frequency mismatch of the cavity resonances affects the spectro-temporal properties of the generated entangled photon pairs in the same way for SPDC and SFWM.
A stronger mismatch reduces the generation rate of photon pairs, 
shortens the auto-correlation time and in theory should lead to temporal modulation of auto-correlation functions of the generated fields. It also distorts spectrum of the generated fields, affects dynamics of the photon pairs generation in the pulsed pump regime.
We provide general expressions for these quantities in the presence of the frequency mismatch.
Also the frequency mismatch affects the temporal mode of a photon heralded upon time-resolved detection of its entangled counterpart.
Results are obtained using an approximation of slowly varying temporal amplitudes of interactivity fields and a more detailed study beyond the approximation can be conducted.

Second, we observe several of the above effects in the experiment on parametric generation of photon pairs in a triply-resonant monolithic nonlinear whispering-gallery mode resonator.
Namely, using temperature tuning of the resonator and pump frequency sweep, we measured the frequency mismatch via two methods and found an agreement between the results within 3 to 20 percent.
We measured the photon count rates as a function of the parametric resonance frequency mismatch.
Furthermore, we investigate the transient response of the parametric fluorescence to the pulsed excitation for increasing values of the frequency mismatch and observed an agreement with the predicted shortening of the rise time of the fluorescence.
The investigation of the influence of the frequency mismatch on the spectrum and auto-correlation function of the generated fields requires spectral resolution on the MHz-scale and high stability of the experiment.

Our results show that the control of the frequency mismatch in the considered systems is important to generate narrowband entangled photon pairs with the desired spectro-temporal properties, and to herald single-photon pulses in a tailored temporal mode.
Furthermore, results present  to the best of our knowledge the first time-resolved study of pulsed dynamics of the cavity-assisted photon pairs generation in SPDC. 
The pulsed regime of the photon pairs generation provides a higher peak power and (in case of SFWM) higher generation efficiency as compared to the continuous regime, timing information about generated photon pairs \cite{Nielsen:09}.
It also has a potential to control the time-energy correlations of the generated photons via the temporal modulation of the pump pulses \cite{Helt:10, Patera2012, Grassani2016,Brecht2016}.
The theoretical model developed in this work will provide a basis to study control of the photon pairs entanglement via the temporally modulated pump pulses and heralding single photons within the tailored temporal mode as it has been already demonstrated in SFWM in atomic systems \cite{Chen2010}.

We anticipate that obtained results can be of interest for a number of groups generating narrowband photon pairs in a cavity-assisted configurations of SPDC and SFWM, particularly, with an application for quantum information protocols in the temporal domain.

\section{Acknowledgements}\label{sec:Acknowledgements}

The authors thank Andrea Aiello, Alexander Otterpohl, Golnoush Shafiee, Ulrich Vogl, Florian Sedlmeir, and Harald G. L. Schwefel for useful discussions.

\appendix
%%%%%%%%%%%%%%%%%%%%%%%%%%%%%%%%%%%
\section{Pump field in a triply resonant configuration}\L{appx:triply}

In a triply resonant configuration the pump filed resonates in the nonlinear optical cavity along with the signal and idler fields (in contrast to the doubly resonant case).
The slowly varying amplitude of the intracavity pump field $\a_\tx{p}(t)$ that drives a spontaneous parametric conversion (SPDC or SFWM) can be calculated as the convolution of the input pump amplitude $\a^\text{in}_\tx{p}(t)$ with the cavity impulse response function $\G_\tx{p}(t)$ \cite{Bader2013}:
	\begin{align}
	\begin{split}
	& \a_\tx{p}(t) = {\sqrt{\g_\tx{p}'}}\int \G_\tx{p}(t-t') \a^\text{in}_\tx{p}(t') \ud t', \\
	& \text{where}\; \G_\tx{p}(t) = \tx{exp}(-(\g_\tx{p}/2 - i\D_\tx{p})t) \theta(t) \, . \L{Gamma_def}
	\end{split}
	\end{align}
Here $\g_\text{p}'$ and $\g_\tx{p}$ represents coupling and total loss rates of the pump field; $\tt(t)$ is the Heaviside step function; $\D_\tx{p} = \w_\tx{p}- \w_\tx{p}^{(0)}$ is a possible frequency detuning of the pump field frequency from the cavity resonance frequency.
The expression for the impulse response function $\G(t)$ in (\ref{Gamma_def}) is obtained within the following approximation for the slowly varying amplitude of the pump field: $\a(t)-\a(t-\t) \approx \t \dot{\a}(t)$, where $\t$ is the cavity round-trip time. Moreover, all theoretical results derived in this paper assume this approximation.

In the following, we consider continuous and pulsed regimes of the cavity excitation.
For a continuous wave pump, the amplitude of the external pump field $\a^\text{in}_\tx{p}(t) = \a^\text{in}_\tx{p}$ is constant. The intra-cavity field oscillates at the frequency of the external pump with the amplitude $\a_\tx{p}$ and with the stored intra-cavity energy E$_\tx{p}$:
	\begin{align}
	& \a_\tx{p} = \frac{\sqrt{\g_\tx{p}'}}{\g_\tx{p}/2-i\D_\tx{p}}  \a^\text{in}_\tx{p}, \label{a_CW}\\
	&\text{E}_\tx{p} \equiv |\a_\tx{p}|^2 = \frac{\g_\tx{p}'}{\g_\tx{p}^2/4+\D_\tx{p}^2} \; \text{P}^\text{in}_\tx{p} , \label{eq:n_p}
	\end{align}
where P$^\text{in}_\tx{p} = |\a^\text{in}_\tx{p}|^2$ is the average input pump power.
The pump intra-cavity energy E$_\tx{p}$ is a Lorentzian function of the detuning. 
Using the input-output relation $\a_\tx{p}^\tx{out} = \sqrt{\g_\tx{p}'} \a_\tx{p} - \a^\text{in}_\tx{p}$, we get the amplitude $\a_\tx{p}^\tx{out}$ of the reflected pump field. 
The power of the reflected field P$_\tx{p}^\text{out} = |\a_\tx{p}^\text{out}|^2$ normalized by the input pump power reads:
	\begin{align}
	& \frac{\text{P}_\tx{p}^\text{out}}{\text{P}^\text{in}_\tx{p}}= 1- \kappa_\tx{p} (1-\kappa_\tx{p}) \frac{\g_\tx{p}^2}{\g_\tx{p}^2/4+\D_\tx{p}^2} \, ,
	\label{eq:Pout}
	\end{align}
where $\kappa_\tx{p} = \g_\tx{p}'/\g_\tx{p}$ gives the coupling coefficient. Thus, the reflected pump power is an inverted Lorentzian function of the frequency detuning $\Delta_\tx{p}$ with the FWHM given by the pump resonance bandwidth $\g_\tx{p}$.

\begin{figure}[h]
\center{\includegraphics[width=0.5\linewidth]{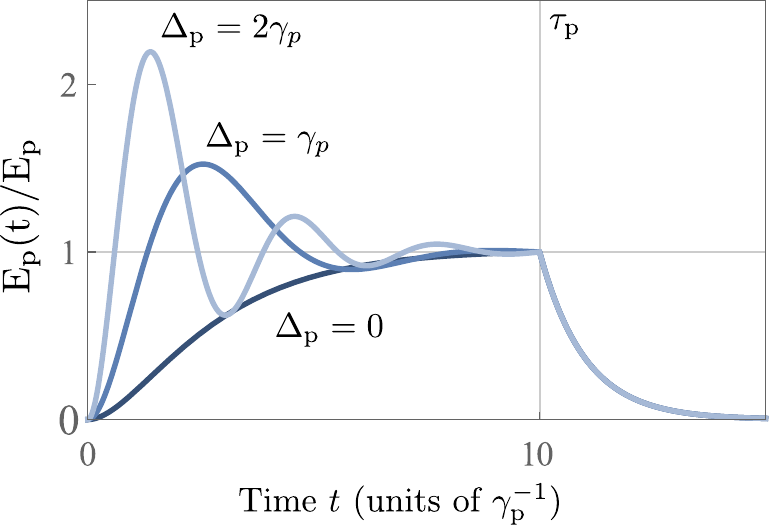}}
	\caption{Time-dependent pump energy E$_\tx{p}(t)$ stored in the optical cavity under the cavity pulsed excitation. The external pump pulse duration is $\tau_\tx{p} = 10 \g^{-1}_\tx{p}$. The graphs are plotted for different values of the pump detuning $\Delta_\tx{p}$. The stored energy  E$_\tx{p}(t)$ is normalized to the steady state value E$_\text{p}$ (see~\ref{eq:n_p}).}
	\label{fig:pump}
	\end{figure}

For a pulsed incident pump, with rectangular temporal shape, with amplitude $\a_\tx{p}^\tx{in}$ and duration $\t_\tx{p}$, the intracavity pump field amplitude reads (according the single-mode approximation (\ref{Gamma_def}))
	\begin{align}
	\a_\tx{p}(t) 
	= \a_\tx{p} \times \left\{\begin{array}{ll}
	f(t), & 0 \leq t \leq \t_p\\
	f(\t_\tx{p}) e^{-\g_\tx{p}(t-\t_\tx{p})/2}, & t>\t_\tx{p}
	\end{array}\right.
	\label{eq:npt}
	\end{align}
where the function $f(t) = 1-e^{-(\g_\tx{p}/2-i\D_\tx{p})t}$ describes the step response of the cavity mode and contains two terms - induced term at the frequency of the external pump and a freely decaying term at the resonance frequency of the cavity.
Fig.~\ref{fig:pump} illustrates the time dependence of the stored energy E$_\tx{p}(t) = |\a_p(t)|^2$ inside the cavity normalized to the steady state value for different values of detuning $\Delta_\tx{p}$. 
The stored pump energy depends  on the absolute value $|\Delta_\tx{p}|$.
At the beginning the dependence shows the rise of the energy accompanied by the ringing effect for nonzero detunings, that can be attributed to the beating between the induced and the free decaying term.
The rise time of the intracavity energy shortens with an increased absolute detuning $|\Delta_\tx{p}|$.
The energy reaches its steady state value given by the (\ref{eq:n_p}). 
At the end of the pulse the energy reduces exponentially with the rate $\g_\tx{p}$ that is defined solely by the effects of intracavity losses and leakage through the coupling mirror and does not depend on the detuning.

%%%%%%%%%%%%%%%%%%%%%%%%%%%%%%%%%%%
\section{Heisenberg-Langevin theory of cavity-assisted SPDC and SFWM}\L{appx:HL}
	
Generation of single-mode signal and idler fields due to parametric down-conversion or four-wave mixing in a nonlinear optical cavity %below the oscillation threshold 
can be described using the following Heisenberg-Langevin equations for slowly varying intracavity amplitudes of signal and idler fields (see, for example, \cite{Fabre1989}):
	\begin{align}
	& \frac{\tx{d}}{\tx{dt}} \hat{a}_\tx{s} (t)=  - \frac{\gamma_\tx{s}}{2} \hat{a}_\tx{s}(t)  +  \frac{g}{2}  \hat{a}_\tx{i}\ssym{2}(t) \a_\tx{p}^{j=1,2}(t) e^{-i\D t}   +  
\sqrt{\g_s'} \hat a_s^\inn(t) + \sqrt{\g_s''} \hat b_s^\inn(t), \L{B1}
 \\
	& \frac{\tx{d} }{\tx{dt}} \hat{a}_\tx{i} (t)=  - \frac{\gamma_\tx{i}}{2} \hat{a}_\tx{i}(t)  +  \frac{g}{2}  \hat{a}_\tx{s}\ssym{2}(t) \a_\tx{p}^{j=1,2}(t) e^{-i\D t} +  
\sqrt{\g_i'} \hat a_i^\inn(t) + \sqrt{\g_i''} \hat b_i^\inn(t).
	\label{eq:rateequationsshort}
	\end{align} 
Modulus squared of intracavity amplitudes give an average number of photons in the corresponding cavity mode.
$\g_\tx{s,i}'$ and $\g_\tx{s,i}''$ are coupling rates and intracavity loss rates of the fields, respectively; $\g_\tx{s,i} = \g_\tx{s,i}' + \g_\tx{s,i}''$ define response rates.
$\a_\tx{p}(t)$ describes a slowly-varying temporal amplitude of the pump field, such that $|\a_\text{p}(t)|^2$ is either equal to the power P$_\text{p}(t)$ of the pump for a doubly resonant configuration or equal to the energy E$_\tx{p}(t)$ stored in the pump mode for a triply resonant configuration. 
Further, $j=1$ in the case of parametric down-conversion and $j=2$  in the case of four-wave mixing.
$g$ characterizes efficiency of a nonlinear process and depends, particularly, on a nonlinear susceptibility of the medium, and on the spatial overlap of interacting modes.
It is measured in 1/s$\cdot\sqrt{\text{W}}$ for a doubly resonant configuration and in 1/s$\cdot\sqrt{\text{J}}$ for a triply resonant configuration.
$\hat a_\tx{s,i}^\inn(t), \hat b_\tx{s,i}^\inn(t)$ are quantized amplitudes of Langevin noise sources associated with the coupling and dissipation, respectively. This noise has zero mean amplitude with the following nonzero correlation functions: 
	\begin{align}
	& \<\hat a_{s,i}^\inn(t) \hat a_{s,i}^{\inn\+}(t')\> = \dd(t-t') \quad \text{and} \quad 
	\<\hat b_{s,i}^\inn(t) \hat b_{s,i}^{\inn\+}(t')\> = \dd(t-t'). \L{crlt}
	\end{align} 
Solution of equations (\ref{B1}) and (\ref{eq:rateequationsshort}) provides to find characteristics of the generated output fields using the following input-output relations 
	\begin{align}
	& \hat a_\text{s,i}^\text{out} = \sqrt{\g_\text{s,i}'} \; \hat a_\text{s,i} - \hat a_\text{s,i}^\text{in}
	\end{align}
Equations (\ref{eq:rateequationsshort}) are linear and their general solution can be written as the following linear transformation of input fields into output fields
	\begin{align}
	& \hat a_{s,i}^{\o} = \bU_{s,i} \; \hat a_{s,i}^{\inn} +  \bu_{s,i} \; \hat b_{s,i}^{\inn} + \bV_{s,i} \; \hat a_{i,s}^{\inn\+}+ \bv_{s,i} \; \hat b_{i,s}^{\inn\+}
	\L{transform}
	\end{align} 
where $\bU, \bV, \bu, \bv$ denote integral transforms. For example, $\bU_\tx{s} \hat a_\tx{s}^\i$ means $\int \ud t' U_\text{s}(t,t') \hat a_s^\inn(t') $ with the integral kernel $U_\tx{s}(t,t')$. 

We assume the regime of weak nonlinear process and solve the linear equations in the first order approximation with respect to $g$. 
The integral kernels then read 
	\begin{align}
	\begin{split}
	& U_{s,i}(t,t') = \g_{s,i}' \G_{s,i}(t-t') - \dd(t-t'),\\
	& u_{s,i}(t,t') = \sqrt{\g_{s,i}' \g_{s,i}''} \G_{s,i}(t-t'),\\
	& V_{s,i}(t,t') = \sqrt{\g_{s,i}' \g_{i,s}'} \; \frac{g}{2} \int \ud t'' \G_{s,i}(t-t'') \a_\tx{p}^{j=1,2}(t'') e^{-i\D t''} \G_{i,s} (t''-t'),\\
	& v_{s,i}(t,t') = \sqrt{\frac{\g_{i,s}''}{\g_{i,s}'}} \; V_{s,i}(t,t')
	\end{split}
	\L{sol}
	\end{align} 
where $\G_{s,i}(t) = \text{exp}(-\g_{s,i}t/2) \theta(t)$ are impulse response functions of the cavity at signal and idler resonances, respectively. The solution is valid for arbitrary slow varying pump, detuning and intracavity losses.

\subsection{General expressions for characteristics of output fields}

Below we give general expressions for main characteristics of the output signal and idler fields using solutions for the Heisenberg-Langevin equations (\ref{sol}) together with the correlation functions of input fields (\ref{crlt}).

First-order nonzero auto- and cross- correlation functions read, respectively:

	\begin{align}
	& G_\tx{ss,ii}^{(1)}(t,t') = \<\hat a_\tx{s,i}^{\text{out}\+}(t) \hat a_\tx{s,i}^\text{out}(t')\> 
	=  \int \ud t'' \[ V_\tx{s,i}^*(t,t'') V_\tx{s,i}(t',t'') + v_\tx{s,i}^*(t,t'') v_\tx{s,i}(t',t'')\], \L{Gssii}\\
	& G_\tx{si}^{(1)}(t,t') = \<\hat a_{s}^\text{out}(t) \hat a_{i}^\text{out}(t')\> = \int \ud t'' \[U_{s}(t,t'') V_i(t',t'') + u_{s}(t,t'') v_i(t',t'')\]
	\end{align} 
Then the output photon flux reads 
	\begin{align}
	& n_\tx{s,i}^\text{out}(t) \equiv \<\hat a_\tx{s,i}^{\text{out}\+}(t) \hat a_\tx{s,i}^\text{out}(t)\> = G_\tx{ss,ii}^{(1)}(t,t)
	\L{nsi_gen}
	\end{align} 
Non-normalized second-order auto- and cross-correlation functions of signal and idler fields can be calculated using the corresponding first-order correlation functions, thanks to the factorization theorem \cite{Bocquillon2009}. The functions read
	\begin{align}
	& G_\tx{ss,ii}^{(2)}(t,t') = \<\hat a_\tx{s,i}^{\text{out}\+}(t') \hat a_\tx{s,i}^{\text{out}\+}(t) \hat a_\tx{s,i}^\text{out}(t) \hat a_\tx{s,i}^\text{out}(t')\>
	= n_\tx{s,i}^\tx{out}(t) n_\tx{s,i}^\tx{out}(t')  + |G_\tx{ss,ii}^{(1)}(t,t')|^2, \\
	& G_\tx{si}^{(2)}(t,t') = \<\hat a_\tx{i}^{\text{out}\+}(t') \hat a_\tx{s}^{\text{out}\+}(t) \hat a_\tx{s}^\text{out}(t) \hat a_\tx{i}^\text{out}(t')\>
	= n_\tx{s}^\tx{out}(t) n_\tx{i}^\tx{out}(t') + |G_\tx{si}^{(1)}(t,t')|^2
	\end{align} 
There are two nonzero contributions to the functions, nameley, the non-correlated and correlated one.
The normalized second-order correlation functions read
	\begin{align}
	& g_\tx{ss,ii}^{(2)}(t,t') = \frac{G_\tx{ss,ii}^{(2)}(t,t') }{n_\tx{s,i}^\tx{out}(t) n_\tx{s,i}^\tx{out}(t')} =1+ \frac{|G_\tx{ss,ii}^{(1)}(t,t')|^2}{n_\tx{s,i}^\tx{out}(t) n_\tx{s,i}^\tx{out}(t')},\\
	& g_\tx{si}^{(2)}(t,t') = \frac{G_\tx{si}^{(2)}(t,t') }{n_\tx{s}^\tx{out}(t) n_\tx{i}^\tx{out}(t')} =1+ \frac{|G_\tx{si}^{(1)}(t,t')|^2}{n_\tx{s}^\tx{out}(t) n_\tx{i}^\tx{out}(t')}.
	\end{align} 

The spectral first-order auto-correlation function of the fields can be calculated as 
	\begin{align}
	& \<\hat a_\tx{s,i}^\+(\w) \hat a_\tx{s,i}(\w')\> = \frac{1}{2\pi} \iint G_\tx{ss,ii}^{(1)}(t,t') e^{-i\w t} e^{i\w' t'} \ud t \ud t'.
	\end{align} 
	%

%%%%%%%%%%%%%%%%%%%%%%%%%%%%%%%%%%%	
\subsection{Continuous wave pump}

Here we calculate properties of the generated output signal and idler fields using the above results.
For a CW pump the amplitude of the pump field is constant $\a_\tx{p}(t)=\a_\tx{p}$ and one can calculate integrals in expressions (\ref{sol}). Substituting integration results into (\ref{Gssii}) one gets first-order auto-correlation functions of signal and idler fields. 
Finally, using (\ref{nsi_gen}) one gets that the output flux of the signal and idler photons is constant and reads:
	\begin{align}
	n_\tx{s,i}^\text{out} = \kappa_\tx{s,i} r_\tx{si}.
	\L{N-si}
	\end{align}
Here $\kappa_\tx{s,i} = \g_\tx{s,i}'/\g_\tx{s,i}$ are the coupling coefficients of the signal and idler fields. $r_\text{si}$ has the meaning of the rate of photon pairs generation and reads
	\begin{align}
	\begin{split}
	&  r_\tx{si} =  \frac{1}{\g_s^{-1} + \g_i^{-1}} \(\frac{\text{P}_\tx{p}}{\text{P}_0}\)^j \frac{1}{1+\D^2/\bar\g^2},  \L{r_def}\\
	& \tx{where} \; j=1 \; (\tx{SPDC}) \quad \tx{or} \quad j=2 \; (\tx{SFWM}).
	\end{split}
	\end{align} 
Here P$_\tx{p} = |\a_\text{p}|^2$ is the power of the field that pumps nonlinear cavity in a doubly resonant configuration;
P$_0$ is the threshold power of the nonlinear oscillation of the system on resonance for considered model. P$_0=\g_\tx{s} \g_\tx{i}/g^2$ for SPDC process and P$_0=\sqrt{\g_\tx{s} \g_\tx{i}/g^2}$ for SFWM process.
An explicit expression for the threshold power is defined by the constant $g$, which depends on a nonlinear process. For example, the expression can be found in \cite{Ilchenko:03}.
$\bar\g = (\g_\tx{s} + \g_\tx{i})/2$ is the average response rate of the optical cavity. 
The normalized cross-correlation function of output signal and idler fields reads:
	\begin{align}
	& g_\tx{si}^{(2)}(\t) = 1 + \frac{1}{r_\tx{si}} \frac{1}{\g_s^{-1} + \g_i^{-1}} \cdot
	\left\{\begin{array}{ll}
	e^{\g_\tx{s} \t}, & \t<0\\
	e^{-\g_\tx{i} \t}, & \t>0
	\end{array}\right. \, .
	\end{align} 
In the limit of low pairs production rate, i.e. $r_\tx{si} \ll 1$, the expression coincides with (\ref{eq:Gsi}).

The auto-correlation functions read
	\begin{align}
	\begin{split}
	& g_\tx{ss,ii}^{(2)}(\t) \\
	&= \frac{(\g_i^2 e^{-\g_s |\t|} + \g_s^2 e^{-\g_i |\t|})(\bar\g^2 + \D^2) - 2\g_s \g_i (\bar\g^2 - \D^2) \cos(\t\cdot \D) e^{-\bar\g|\t|} + 4 \g_s \g_i \bar \g \D \sin(|\t| \cdot \D) e^{-\bar\g|\t|}}{\bar\g^2 ((\g_s-\g_i)^2+4\D^2)}
	\L{g2_general}
	\end{split}
	\end{align} 
In the limit $\g_s = \g_i = \g$ the expression simplifies to (\ref{eq:gssii}).

Spectra of signal and idler fields defined as $\<\hat a_\tx{s,i}^\+(\w) \hat a_\tx{s,i}(\w')\> = S_\tx{s,i}(\w)\dd(\w-\w')$ read
	\begin{align}
	& S_\tx{s,i}(\w) =  4\kappa_\tx{s,i} \(\frac{\text{P}_\tx{p}}{\text{P}_0}\)^{j=1,2}\frac{1}{\[1+4\w^2/\g_\tx{s,i}^2\]\[1+4(\w-\D)^2/\g_\tx{i,s}^2\]}
	\end{align} 

\bibliographystyle{unsrt}
\bibliography{biblio_VA,MyCollection}

\begin{thebibliography}{10}

\bibitem{Kimble2008}
H.~J. Kimble.
\newblock {The quantum internet}.
\newblock {\em Nature}, 453(7198):1023--1030, jun 2008.

\bibitem{Lvovsky2009}
Alexander~I. Lvovsky, Barry~C. Sanders, and Wolfgang Tittel.
\newblock {Optical quantum memory}.
\newblock {\em Nature Photonics}, 3(12):706--714, dec 2009.

\bibitem{Simon2010}
C.~Simon, M.~Afzelius, J.~Appel, A.~Boyer de~la Giroday, S.~J. Dewhurst,
  N.~Gisin, C.~Y. Hu, F.~Jelezko, S.~Kr{\"o}ll, J.~H. M{\"u}ller, J.~Nunn,
  E.~S. Polzik, J.~G. Rarity, H.~De~Riedmatten, W.~Rosenfeld, A.~J. Shields,
  N.~Sk{\"o}ld, R.~M. Stevenson, R.~Thew, I.~A. Walmsley, M.~C. Weber,
  H.~Weinfurter, J.~Wrachtrup, and R.~J. Young.
\newblock Quantum memories.
\newblock {\em The European Physical Journal D}, 58(1):1--22, May 2010.

\bibitem{Specht2011}
Holger~P. Specht, Christian N{\"{o}}lleke, Andreas Reiserer, Manuel Uphoff,
  Eden Figueroa, Stephan Ritter, and Gerhard Rempe.
\newblock {A single-atom quantum memory}.
\newblock {\em Nature}, 473(7346):190--193, may 2011.

\bibitem{Sangouard2011}
Nicolas Sangouard, Christoph Simon, Hugues de~Riedmatten, and Nicolas Gisin.
\newblock Quantum repeaters based on atomic ensembles and linear optics.
\newblock {\em Rev. Mod. Phys.}, 83:33--80, Mar 2011.

\bibitem{Fekete2013}
Julia Fekete, Daniel Riel\"ander, Matteo Cristiani, and Hugues de~Riedmatten.
\newblock Ultranarrow-band photon-pair source compatible with solid state
  quantum memories and telecommunication networks.
\newblock {\em Phys. Rev. Lett.}, 110:220502, May 2013.

\bibitem{Neergaard-Nielsen:07}
J.~S. Neergaard-Nielsen, B.~Melholt Nielsen, H.~Takahashi, A.~I. Vistnes, and
  E.~S. Polzik.
\newblock High purity bright single photon source.
\newblock {\em Opt. Express}, 15(13):7940--7949, Jun 2007.

\bibitem{Scholz2009}
Matthias Scholz, Lars Koch, and Oliver Benson.
\newblock Statistics of narrow-band single photons for quantum memories
  generated by ultrabright cavity-enhanced parametric down-conversion.
\newblock {\em Phys. Rev. Lett.}, 102:063603, Feb 2009.

\bibitem{Spring:13}
Justin~B. Spring, Patrick~S. Salter, Benjamin~J. Metcalf, Peter~C. Humphreys,
  Merritt Moore, Nicholas Thomas-Peter, Marco Barbieri, Xian-Min Jin, Nathan~K.
  Langford, W.~Steven Kolthammer, Martin~J. Booth, and Ian~A. Walmsley.
\newblock On-chip low loss heralded source of pure single photons.
\newblock {\em Opt. Express}, 21(11):13522--13532, Jun 2013.

\bibitem{Averchenko2016}
Valentin Averchenko, Denis Sych, Gerhard Schunk, Ulrich Vogl, Christoph
  Marquardt, and Gerd Leuchs.
\newblock {Temporal shaping of single photons enabled by entanglement}.
\newblock {\em Physical Review A}, 96(4):043822, oct 2017.

\bibitem{Ou1999}
Z~Ou and Y~Lu.
\newblock {Cavity Enhanced Spontaneous Parametric Down-Conversion for the
  Prolongation of Correlation Time between Conjugate Photons}.
\newblock {\em Physical Review Letters}, 83(13):2556--2559, sep 1999.

\bibitem{Fortsch2013}
M~F{\"{o}}rtsch, J~U F{\"{u}}rst, C~Wittmann, D~Strekalov, A~Aiello,
  M~Chekhova, Christine Silberhorn, Gerd Leuchs, and Christoph Marquardt.
\newblock {A versatile source of single photons for quantum information
  processing.}
\newblock {\em Nature communications}, 4:1818, 2013.

\bibitem{KHLuo2015}
Kai-Hong Luo, Harald Herrmann, Stephan Krapick, Benjamin Brecht, Raimund
  Ricken, Viktor Quiring, Hubertus Suche, Wolfgang Sohler, and Christine
  Silberhorn.
\newblock Direct generation of genuine single-longitudinal-mode narrowband
  photon pairs.
\newblock {\em New Journal of Physics}, 17(7):073039, 2015.

\bibitem{Matsko2005ReviewOA}
A.~B. Matsko, A.~A. Savchenkov, Dmitry~V. Strekalov, V.~S. Ilchenko, and Lili
  Maleki.
\newblock Review of applications of whispering-gallery mode resonators in
  photonics and nonlinear optics.
\newblock 2005.

\bibitem{Strekalov2016}
Dmitry~V Strekalov, Christoph Marquardt, Andrey~B Matsko, Harald G~L Schwefel,
  and Gerd Leuchs.
\newblock {Nonlinear and quantum optics with whispering gallery resonators}.
\newblock {\em Journal of Optics}, 18(12):123002, dec 2016.

\bibitem{Grassani:15}
Davide Grassani, Stefano Azzini, Marco Liscidini, Matteo Galli, Michael~J.
  Strain, Marc Sorel, J.~E. Sipe, and Daniele Bajoni.
\newblock Micrometer-scale integrated silicon source of time-energy entangled
  photons.
\newblock {\em Optica}, 2(2):88--94, Feb 2015.

\bibitem{Lu2016}
Xiyuan Lu, Wei~C. Jiang, Jidong Zhang, and Qiang Lin.
\newblock {Biphoton Statistics of Quantum Light Generated on a Silicon Chip}.
\newblock {\em ACS Photonics}, 3(9):1626--1636, sep 2016.

\bibitem{Eckardt:91}
Robert~C. Eckardt, C.~D. Nabors, William~J. Kozlovsky, and Robert~L. Byer.
\newblock Optical parametric oscillator frequency tuning and control.
\newblock {\em J. Opt. Soc. Am. B}, 8(3):646--667, Mar 1991.

\bibitem{Breunig2014}
Ingo Breunig, Anni B{\"{u}}ckle, Christoph~S. Werner, and Karsten Buse.
\newblock {Non-Lorentzian pump resonances in whispering gallery optical
  parametric oscillators}.
\newblock volume 8960, page 896007, mar 2014.

\bibitem{Tang1995}
C.~L. Tang and L.~K. Cheng.
\newblock {\em {Fundamentals of Optical Parametric Processes and
  Oscillations}}.
\newblock Harwood Academic Press, 1 edition, 1995.

\bibitem{Zielinska2014}
Joanna~A. Zieli{\'{n}}ska and Morgan~W. Mitchell.
\newblock {Theory of high gain cavity-enhanced spontaneous parametric
  down-conversion}.
\newblock {\em Physical Review A}, 90(6):063833, dec 2014.

\bibitem{Haase:09}
Albrecht Haase, Nicolas Piro, J\"{u}rgen Eschner, and Morgan~W. Mitchell.
\newblock Tunable narrowband entangled photon pair source for resonant
  single-photon single-atom interaction.
\newblock {\em Opt. Lett.}, 34(1):55--57, Jan 2009.

\bibitem{Fortsch2014}
Michael F\"ortsch, Gerhard Schunk, Josef~U. F\"urst, Dmitry Strekalov, Thomas
  Gerrits, Martin~J. Stevens, Florian Sedlmeir, Harald G.~L. Schwefel, Sae~Woo
  Nam, Gerd Leuchs, and Christoph Marquardt.
\newblock Highly efficient generation of single-mode photon pairs from a
  crystalline whispering-gallery-mode resonator source.
\newblock {\em Phys. Rev. A}, 91:023812, Feb 2015.

\bibitem{Heebner2004}
John Heebner, Vincent Wong, Aaron Schweinsberg, Robert Boyd, and Deborah
  Jackson.
\newblock Optical transmission characteristics of fiber ring resonators.
\newblock 40:726 -- 730, 07 2004.

\bibitem{Garay-Palmett2013}
K~Garay-Palmett, Y~Jeronimo-Moreno, and A~B U'Ren.
\newblock {Theory of cavity-enhanced spontaneous four wave mixing}.
\newblock {\em Laser Physics}, 23(1):015201, jan 2013.

\bibitem{Vernon2017}
Z.~Vernon, M.~Menotti, C.~C. Tison, J.~A. Steidle, M.~L. Fanto, P.~M. Thomas,
  S.~F. Preble, A.~M. Smith, P.~M. Alsing, M.~Liscidini, and J.~E. Sipe.
\newblock {Truly unentangled photon pairs without spectral filtering}.
\newblock {\em Optics Letters}, 42(18):3638, sep 2017.

\bibitem{Weiss1995}
D.~S. Weiss, V.~Sandoghdar, J.~Hare, V.~Lef{\`{e}}vre-Seguin, J.-M. Raimond,
  and S.~Haroche.
\newblock {Splitting of high-Q Mie modes induced by light backscattering in
  silica microspheres}.
\newblock {\em Optics Letters}, 20(18):1835, sep 1995.

\bibitem{Little1997}
Brent~E. Little, Juha-Pekka Laine, and Sai~T. Chu.
\newblock {Surface-roughness-induced contradirectional coupling in ring and
  disk resonators}.
\newblock {\em Optics Letters}, 22(1):4, jan 1997.

\bibitem{MacLean2018}
Jean-Philippe~W. MacLean, John~M. Donohue, and Kevin~J. Resch.
\newblock Direct characterization of ultrafast energy-time entangled photon
  pairs.
\newblock {\em Phys. Rev. Lett.}, 120:053601, Jan 2018.

\bibitem{Loudon2000}
R~Loudon.
\newblock {\em {The Quantum Theory of Light}}.
\newblock Oxford University Press, 2000.

\bibitem{Drummond1990}
P.~D. Drummond and M.~D. Reid.
\newblock {Correlations in nondegenerate parametric oscillation. II. Below
  threshold results}.
\newblock {\em Physical Review A}, 41(7):3930--3949, apr 1990.

\bibitem{Bocquillon2009}
E.~Bocquillon, C.~Couteau, M.~Razavi, R.~Laflamme, and G.~Weihs.
\newblock {Coherence measures for heralded single-photon sources}.
\newblock {\em Physical Review A}, 79(3):035801, mar 2009.

\bibitem{McNeil1983}
K.~J. McNeil and C.~W. Gardiner.
\newblock Quantum statistics of parametric oscillation.
\newblock {\em Phys. Rev. A}, 28:1560--1566, Sep 1983.

\bibitem{Luo2015}
Kai-Hong Luo, Harald Herrmann, Stephan Krapick, Benjamin Brecht, Raimund
  Ricken, Viktor Quiring, Hubertus Suche, Wolfgang Sohler, and Christine
  Silberhorn.
\newblock {Direct generation of genuine single-longitudinal-mode narrowband
  photon pairs}.
\newblock {\em New Journal of Physics}, 17(7):073039, aug 2015.

\bibitem{Averchenko2011}
V.~A. Averchenko, Yu.~M. Golubev, C.~Fabre, and N.~Treps.
\newblock Quantum correlations and fluctuations in the pulsed light produced by
  a synchronously pumped optical parametricoscillator below its oscillation
  threshold.
\newblock {\em The European Physical Journal D}, 61(1):207--214, Jan 2011.

\bibitem{Mosley2008}
Peter~J. Mosley, Jeff~S. Lundeen, Brian~J. Smith, Piotr Wasylczyk, Alfred~B.
  U'Ren, Christine Silberhorn, and Ian~A. Walmsley.
\newblock Heralded generation of ultrafast single photons in pure quantum
  states.
\newblock {\em Phys. Rev. Lett.}, 100:133601, Apr 2008.

\bibitem{Furst2010natural}
J.~F{\"{u}}rst, D.~V. Strekalov, D.~Elser, M.~Lassen, U.~L. Andersen, Ch.
  Marquardt, and G.~Leuchs.
\newblock {Naturally Phase-Matched Second-Harmonic Generation in a
  Whispering-Gallery-Mode Resonator}.
\newblock {\em Physical Review Letters}, 104(15):153901, 2010.

\bibitem{Furst2010}
J.~F{\"{u}}rst, D.~Strekalov, D.~Elser, A.~Aiello, U.~Andersen, Ch. Marquardt,
  and G.~Leuchs.
\newblock {Low-Threshold Optical Parametric Oscillations in a Whispering
  Gallery Mode Resonator}.
\newblock {\em Physical Review Letters}, 105(26):2--5, dec 2010.

\bibitem{Michael2013}
Michael F{\"{o}}rtsch, Josef~U F{\"{u}}rst, Christoffer Wittmann, Dmitry
  Strekalov, Andrea Aiello, Maria~V Chekhova, Christine Silberhorn, Gerd
  Leuchs, and Christoph Marquardt.
\newblock {A versatile source of single photons for quantum information
  processing}.
\newblock {\em Nature Communications}, 4(2):1818, may 2013.

\bibitem{Schunk2015a}
Gerhard Schunk, Ulrich Vogl, Dmitry~V. Strekalov, Michael F{\"{o}}rtsch,
  Florian Sedlmeir, Harald G.~L. Schwefel, Manuela G{\"{o}}belt, Silke
  Christiansen, Gerd Leuchs, and Christoph Marquardt.
\newblock {Interfacing transitions of different alkali atoms and telecom bands
  using one narrowband photon pair source}.
\newblock {\em Optica}, 2(9):773, sep 2015.

\bibitem{Schunk2016}
G.~Schunk, U.~Vogl, F.~Sedlmeir, D.~V. Strekalov, A.~Otterpohl, V.~Averchenko,
  H.~G.~L. Schwefel, G.~Leuchs, and Ch. Marquardt.
\newblock {Frequency tuning of single photons from a whispering-gallery mode
  resonator to MHz-wide transitions}.
\newblock {\em Journal of Modern Optics}, 63(20):2058--2073, nov 2016.

\bibitem{SedlmeirSelective2016}
Florian Sedlmeir, Matthew~R. Foreman, Ulrich Vogl, Richard Zeltner, Gerhard
  Schunk, Dmitry~V. Strekalov, Christoph Marquardt, Gerd Leuchs, and Harald
  G.~L. Schwefel.
\newblock {Polarization-Selective Out-Coupling of Whispering-Gallery Modes}.
\newblock {\em Physical Review Applied}, 7(2):024029, feb 2017.

\bibitem{Breunig2016}
Ingo Breunig.
\newblock {Three-wave mixing in whispering gallery resonators}.
\newblock {\em Laser {\&} Photonics Reviews}, 10(4):569--587, jul 2016.

\bibitem{Strekalov2016jop}
Dmitry~V Strekalov, Christoph Marquardt, Andrey~B Matsko, Harald G~L Schwefel,
  and Gerd Leuchs.
\newblock {Nonlinear and quantum optics with whispering gallery resonators}.
\newblock {\em Journal of Optics}, 18(12):123002, dec 2016.

\bibitem{Schunk2014a}
Gerhard Schunk, Josef~U. F{\"{u}}rst, Michael F{\"{o}}rtsch, Dmitry~V.
  Strekalov, Ulrich Vogl, Florian Sedlmeir, Harald G.~L. Schwefel, Gerd Leuchs,
  and Christoph Marquardt.
\newblock {Identifying modes of large whispering-gallery mode resonators from
  the spectrum and emission pattern}.
\newblock {\em Optics Express}, 22(25):30795, dec 2014.

\bibitem{Michael2014}
Michael F{\"{o}}rtsch, Gerhard Schunk, Josef~U. F{\"{u}}rst, Dmitry Strekalov,
  Thomas Gerrits, Martin~J. Stevens, Florian Sedlmeir, Harald G.~L. Schwefel,
  Sae~Woo Nam, Gerd Leuchs, and Christoph Marquardt.
\newblock {Highly efficient generation of single-mode photon pairs from a
  crystalline whispering-gallery-mode resonator source}.
\newblock {\em Physical Review A}, 91(2):023812, feb 2015.

\bibitem{Breunig2013b}
I~Breunig, B~Sturman, A~B{\"{u}}ckle, C~S Werner, and K~Buse.
\newblock {Structure of pump resonances during optical parametric oscillation
  in whispering gallery resonators}.
\newblock {\em Optics letters}, 38(17):3316--8, sep 2013.

\bibitem{Schlarb1994}
U~Schlarb and K~Betzler.
\newblock {Influence of the defect structure on the refractive indices of
  undoped and Mg-doped lithium niobate}.
\newblock {\em Physical Review B}, 50(2):751--757, 1994.

\bibitem{Weis1985}
R~S Weis and T~K Gaylord.
\newblock {Lithium niobate: Summary of physical properties and crystal
  structure}.
\newblock {\em Applied Physics A Solids and Surfaces}, 37(4):191--203, aug
  1985.

\bibitem{Du2015}
Shengwang Du.
\newblock {Quantum-state purity of heralded single photons produced from
  frequency-anticorrelated biphotons}.
\newblock {\em Physical Review A}, 92(4):043836, oct 2015.

\bibitem{Sych2015}
Denis Sych, Valentin Averchenko, and Gerd Leuchs.
\newblock {Generic method for lossless generation of arbitrarily shaped
  photons}.
\newblock {\em Physical Review A}, 96(5):053847, nov 2017.

\bibitem{Nielsen:09}
B.~Melholt Nielsen, J.~S. Neergaard-Nielsen, and E.~S. Polzik.
\newblock Time gating of heralded single photons for atomic memories.
\newblock {\em Opt. Lett.}, 34(24):3872--3874, Dec 2009.

\bibitem{Helt:10}
L.~G. Helt, Zhenshan Yang, Marco Liscidini, and J.~E. Sipe.
\newblock Spontaneous four-wave mixing in microring resonators.
\newblock {\em Opt. Lett.}, 35(18):3006--3008, Sep 2010.

\bibitem{Patera2012}
G.~Patera, C.~Navarrete-Benlloch, G.J. Valc{\'{a}}rcel, and C.~Fabre.
\newblock {Quantum coherent control of highly multipartite continuous-variable
  entangled states by tailoring parametric interactions}.
\newblock {\em The European Physical Journal D}, 66(9):241, sep 2012.

\bibitem{Grassani2016}
Davide Grassani, Angelica Simbula, Stefano Pirotta, Matteo Galli, Matteo
  Menotti, Nicholas~C. Harris, Tom Baehr-Jones, Michael Hochberg, Christophe
  Galland, Marco Liscidini, and Daniele Bajoni.
\newblock {Energy correlations of photon pairs generated by a silicon microring
  resonator probed by Stimulated Four Wave Mixing}.
\newblock {\em Scientific Reports}, 6(1):23564, jul 2016.

\bibitem{Brecht2016}
Benjamin Brecht, Kai-Hong Luo, Harald Herrmann, and Christine Silberhorn.
\newblock A versatile design for resonant guided-wave parametric
  down-conversion sources for quantum repeaters.
\newblock {\em Applied Physics B}, 122(5):116, Apr 2016.

\bibitem{Chen2010}
J.~F. Chen, Shanchao Zhang, Hui Yan, M.~M.~T. Loy, G.~K.~L. Wong, and Shengwang
  Du.
\newblock Shaping biphoton temporal waveforms with modulated classical fields.
\newblock {\em Phys. Rev. Lett.}, 104:183604, May 2010.

\bibitem{Bader2013}
M.~Bader, S.~Heugel, A.L. Chekhov, M.~Sondermann, and G.~Leuchs.
\newblock {Efficient coupling to an optical resonator by exploiting
  time-reversal symmetry}.
\newblock {\em New Journal of Physics}, 15(12):123008, dec 2013.

\bibitem{Fabre1989}
C.~Fabre, E.~Giacobino, A.~Heidmann, and S.~Reynaud.
\newblock {Noise characteristics of a non-degenerate Optical Parametric
  Oscillator - Application to quantum noise reduction}.
\newblock {\em Journal de Physique}, 50(10):1209--1225, 1989.

\bibitem{Ilchenko:03}
Vladimir~S. Ilchenko, Andrey~B. Matsko, Anatoliy~A. Savchenkov, and Lute
  Maleki.
\newblock Low-threshold parametric nonlinear optics with quasi-phase-matched
  whispering-gallery modes.
\newblock {\em J. Opt. Soc. Am. B}, 20(6):1304--1308, Jun 2003.

\end{thebibliography}

\end{document}